\documentclass[11pt,a4paper]{article}
\usepackage{hyphenat}

\usepackage[utf8]{inputenc}
\usepackage[T1]{fontenc}
\usepackage{amsmath,amssymb,amsfonts} % Para las ecuaciones
\usepackage{graphicx}                % Para las imágenes
\usepackage{booktabs,longtable}      % Para tablas limpias
\usepackage{microtype}               % Mejora la tipografía básica
\usepackage[hidelinks]{hyperref}     % Enlaces limpios para emails/URLs
\newcommand{\short}{\text{short}}
\newcommand{\longg}{\text{long}}
\usepackage[margin=1in]{geometry}

\usepackage[numbers,sort&compress]{natbib} % Citas numéricas ordenadas y comprimidas, ej: [1-3]
\title{Attosecond Path Qubits in High-Harmonic Generation:\\ Classical Dephasing and Trace-Out Decoherence}
\author{
    A.~Marchisio$^{1,2, \dagger}$, 
    C.~Granados$^{3}$, 
    M. F.~Ciappina$^{1,2,4}$, 
    O.~Cohen$^{1,2,5,*}$
}
\date{\small} % Deja la fecha vacía o pon la actual

\begin{document}

\maketitle

% --- Afiliaciones (Formato limpio) ---
\begin{center}
    \small
    $^{1}$Department of Physics, Guangdong Technion - Israel Institute of Technology, Shantou \& 515063, People's Republic of China.\\
    $^{2}$Technion -- Israel Institute of Technology, Haifa \& 32000, Israel.\\
    $^{3}$Eastern Institute of Technology, Ningbo \& 315000, People's Republic of China.\\
    $^{4}$Guangdong Provincial Key Laboratory of Materials and Technologies for Energy Conversion, Guangdong Technion - Israel Institute of Technology, Shantou \& 515063, People's Republic of China.\\
    $^{5}$Solid State Institute, Physics Department and Helen Diller Quantum Center, Technion-Israel Institute of Technology, Haifa \& 3200003, Israel.\\[1ex]
    
    $^{\dagger}$Email: marchisio.andres@gtiit.edu.cn \quad $^{*}$Email: oren@technion.ac.il
\end{center}

\vspace{2em}

% --- Resumen (Abstract estándar) ---
\begin{abstract}
High-harmonic generation (HHG) is governed by interference between electron trajectories. We propose that the dominant short and long trajectories define an experimentally addressable two-level subsystem: an attosecond path qubit (APQ). We formulate a trajectory-resolved density matrix to identify two distinct coherence-loss mechanisms: classical dephasing from ensemble averaging and quantum decoherence arising from the trace-out of unobserved degrees of freedom. By investigating shot-to-shot fluctuations and unresolved transverse momentum, we demonstrate that while dephasing suppresses coherence through averaging, the ``trace-out'' channel produces mixed states even for fixed driving parameters. We explore how these mechanisms modify APQ purity and show that mode selection and conditioning provide operational routes to isolate them. These results establish a reduced-state framework for diagnosing coherence loss in HHG and for engineering trajectory-based quantum states in attosecond interferometry.
\end{abstract}

\vspace{2em}

\section{INTRODUCTION}

Attosecond science has developed rapidly, largely driven by advances in ultrafast laser technology and the ability to generate and characterize light pulses with sub-femtosecond duration \cite{Krausz2009}. These tools
provide direct access to electronic motion on its native time scale and have enabled studies of ultrafast dynamics in atoms, molecules, and condensed-matter systems \cite{Krausz2009}. A central mechanism underlying these developments is high-harmonic generation (HHG), a highly nonlinear parametric process in which an intense laser field upconverts infrared or visible radiation into coherent extreme-ultraviolet emissions. With the continued development of experimental techniques, HHG has been demonstrated in a broad range of physical systems, including atomic and molecular gases, solids, and, more recently, liquids \cite{Li2020,LHuillier1991,Heide2024,Ghimire2019,Ghimire2026}.

Since the early 1990s, the physical mechanism responsible for HHG in gases has been successfully interpreted within the semiclassical three-step model \cite{Kulander1993,Corkum1993}. In this picture, the electron dynamics proceed through tunnel ionization, acceleration in the continuum, and recombination with the parent ion, releasing its energy as high-harmonic photons. The resulting emission spectra exhibit a characteristic non-perturbative structure: an initial rapid decay for the lowest harmonic orders, followed by a broad intensity plateau where harmonic strength remains largely constant, and finally a sharp cutoff region marking the maximum energy achievable by the emitted radiation \cite{Ghimire2026}.

A more detailed description emerged from the strong-field approximation (SFA) in which the HHG emission amplitude is expressed as a coherent sum over saddle-point solutions of the semiclassical action \cite{Lewenstein1994,Amini2019}. These saddle-points correspond to different quantum trajectories (or quantum orbits) followed by the electron in the continuum. Within this framework, two dominant trajectories contribute to the HHG emission: the short and long trajectories, which are distinguished by electron excursion times and exhibit different phase accumulations \cite{Lewenstein1995}, intensity gradient phases \cite{Salires2001}, divergence angles \cite{Lloyd2016,Bellini1998}, and coherence length \cite{Bellini1998,Lynga1999}

Coherent contribution of the short and long quantum trajectories to the measured HHG signal often yields observed interference patterns. These interference signatures typically manifest as spectral and spatial modulations, as demonstrated in early experimental studies of the emission process \cite{Bellini1998,Lynga1999}. Because each trajectory accumulates a different intensity-dependent phase, their interference produces characteristic signatures in both the harmonic spectra and spatial profiles. Subsequent works demonstrated that these contributions can be experimentally separated and controlled through phase-matching conditions, focusing geometry, and spatial filtering \cite{Brugnera2011,Winterfeldt2008}. Spatially resolved measurements have shown that short and long trajectories leave distinct fingerprints in the harmonics at the far-field, enabling their identification and partial separation \cite{Zhang2008,Schapper2010}.

These studies establish that quantum-path interference in HHG is a directly measurable phenomenon encoded in experimentally accessible observables. Yet, despite this level of control, short--long interference has been treated primarily through its macroscopic signatures, such as spectral modulations induced by variations of laser intensity or phase-matching conditions \cite{Zair2008}. More recently, it has been treated through phase-sensitive attosecond interferometric measurements of transient strong-field dynamics \cite{Kneller2025}. In this view, the interference serves mainly as a diagnostic of emission dynamics, rather than as evidence of a coherent two-path quantum state of the electron in the continuum. As a result, despite extensive control over short-long interference, there is still no reduced-state framework that treats the trajectory pair itself as an electronic quantum subsystem. In particular, existing descriptions do not provide a trajectory-basis density matrix from which one can directly quantify path coherence, state purity, distinguishability, or the loss of information to unobserved degrees of freedom. Establishing such a framework is particularly compelling in HHG, where electronic motion unfolds on attosecond timescales and where decoherence, if present, may occur at the shortest natural timescale available in light-matter interaction.

Short-long interference in HHG is closely analogous to a two-path interferometer, where the electron can follow different quantum trajectories before recombining. The resulting harmonic emission therefore encodes both the relative amplitude and phase between these paths, providing a direct probe of the underlying electron dynamics in the continuum. This two-path interferometer interpretation is closely related to the concept of path qubits, extensively studied in the context of decoherence \cite{Englert1996,Drr1998}, where distinct physical trajectories form a controllable Hilbert space. Inspired on the experimentally established picture of short and long trajectories in HHG, we introduce a trajectory-based framework in which these quantum paths define a controllable two-level subsystem (TLS). The relative amplitude and phase of the trajectories can be tuned through experimental parameters such as laser intensity, phase matching, and focusing geometry. The coherent superposition of electron trajectories naturally maps onto a measurement-defined Hilbert space, allowing familiar interferometric observables, such as spectral and spatial modulations, to be interpreted in terms of the TLS dynamics.

Here, we introduce the concept of an attosecond-path qubit (APQ), defined by the coherent superposition of short and long electron trajectories. Utilizing a trajectory-resolved density matrix, we reinterpret strong-field dynamics through the lens of quantum information science, employing Bloch vectors and decoherence channels to quantify the fidelity of the electron's sub-cycle motion. Our approach captures both the coherent evolution of the electronic subsystem and its coupling to environment-like degrees of freedom, allowing us to distinguish between two general mechanisms of coherence loss: classical dephasing from ensemble averaging and intrinsic quantum decoherence arising from the trace-out of unobserved variables. We investigate representative case studies for each, specifically shot-to-shot laser fluctuations and unresolved electron transverse momentum, and demonstrate how operational routes such as conditioning and mode selection provide a means to separate and quantify these channels. This perspective enables us to categorize the complexity of the HHG process not only by its spectral yield, but by the robustness of the encoded quantum information. This framework establishes HHG as a versatile platform for investigating information loss and provides a diagnostic foundation for engineering trajectory-defined quantum states at the sub-cycle timescale of electronic motion. We further identify future directions opened by this reduced-state description, including higher-dimensional trajectory encodings and extensions to quantum
driving fields and multi-mode HHG.

\section{ATTOSECOND-PATH QUBIT}

In strong-field processes, the active electronic contribution is formally treated as a subsystem defined within its respective Hilbert space, \(\mathcal{H}_{e^{-}}\). In HHG, the actual Hilbert space includes the electron, the nuclei, light, and possibly more degrees of freedom that participate in the process. We consider here only a single active electron, and approximate the complete Hilbert space as

\begin{equation}
    \mathcal{H}_{\text{total}} \approx \mathcal{H}_{e^{-}} \otimes \mathcal{H}_{\text{light}} \otimes \mathcal{H}_{\text{other}},
\end{equation}

where \(\mathcal{H}_{\text{light}}\) and \(\mathcal{H}_{\text{other}}\) denote the Hilbert spaces associated with the light subsystem and with any other possible subsystem, respectively. In the standard SFA treatment, both the driving and emitted light fields are typically described as classical entities. If one further assumes that the electron is the only relevant degree of freedom, the SFA provides an excellent qualitative and quantitative model of the HHG process \cite{Salires2001}. Under these assumptions, the system remains effectively
closed, and the dynamics are governed solely by the electron's interaction with the external classical fields \cite{Amini2019}.

The starting point of the SFA is to propose the following ansatz for the electron state~\cite{Lewenstein1994}

\begin{equation}
    |e\rangle = e^{iI_{p}t}|\phi_{0}\rangle + \int d\mathbf{v}b(\mathbf{v},t)|\mathbf{v}\rangle,
\end{equation}

where \(|\phi_{0}\rangle\) is the bound initial electron state, \(\left| \mathbf{v} \right\rangle\) is the continuum state with kinetic energy \(\mathbf{v}\ ,I_{p} = - E_{0}\) is the ionization energy, and \(b(\mathbf{v},t)\) are the probability amplitudes of the electron continuum states.\footnote{We use atomic units, \(m = e = \hslash = 1\), \(\mathbf{v} = \mathbf{p}\) for the electron.} Note that in Eq. (2) we have neglected the ground state depletion (see Supplementary Material (SM) for more details).

As detailed in the SM, HHG emission in the plateau region is primarily governed by the interference of two dominant quantum-orbit contributions: the short and long trajectories. Within the SFA framework, the electron dynamics in the continuum are described by a superposition of Volkov states, which represents the evolution of a free electron driven solely by the laser field (see Eq. (S10) in the SM). By neglecting the residual influence of the ionic core upon ionization, the continuum wavefunction is decomposed into probability amplitudes \(b(\mathbf{v},t)\) associated with these laser-dressed states. By applying the saddle-point approximation to the underlying time and momentum integrals, the continuous evolution collapses into a discrete set of quantum paths. Each trajectory is uniquely characterized by a complex pair of ionization and recombination times \((t_{i},t_{r})\) and a corresponding saddle-point canonical momentum \(\mathbf{p}_{s}(t_{i},t_{r})\). In the deep plateau region, these trajectories map onto distinct, non-overlapping momentum domains, \(V_{short}\) and \(V_{long}\) whose negligible overlap allows for the construction of an approximately orthogonal basis, defining the short and long branches. The existence of these momentum domains provides a natural framework for partitioning the continuum into a reduced Hilbert subspace spanned by the short and long trajectory contributions

\begin{equation}
    \begin{aligned}
        |\short(t)\rangle & \equiv \int_{V_{\short}}^{}d^{3}\mathbf{p}b(\mathbf{p} + \mathbf{A}(t),t)\left| \chi_{\mathbf{p}}(t) \right\rangle, \\
    \quad|\text{long}(t)\rangle & \equiv \int_{V_{\longg}}^{}d^{3}\mathbf{p}b(\mathbf{p} + \mathbf{A}(t),t)\left| \chi_{\mathbf{p}}(t) \right\rangle,
    \end{aligned}
\end{equation}

where the \(V_{\short}\) and \(V_{\longg}\) labels in the integral indicate the integration volume according to the saddle-point solutions (see Eq.~(S11) in the SM). This basis is approximately orthogonal within the plateau region, where short and long trajectories possess distinct, non-coincident canonical momenta. These conditions fail both near the cutoff, where trajectories coalesce and become indistinguishable, and at low energies, \(I_{p} \lesssim \omega\ ,\) where the SFA breaks down. Consequently, the APQ framework is formally restricted to the spectral plateau.

Figure 1(a) illustrates the conceptual framework, showing a two-level system description of a typical HHG experiment. Here, the short-long basis is not merely a theoretical construct but a direct consequence of the distinct electronic paths in the continuum. The pump laser interacts with the gas jet, driving the electron through two dominant quantum trajectories. These paths act as the two arms of an interferometer, whose output (the high-harmonic radiation) carries the phase information of each branch.

With the states presented in Eq.~(3) we define the projectors onto the subspaces corresponding to short and long trajectories

\begin{equation}
    {\hat{P}}_{s} \equiv |\short(t)\rangle\langle \short(t)|,\quad{\hat{P}}_{l} \equiv |\longg(t)\rangle\langle \longg(t)|,
\end{equation}

and we can thus define an effective two-level Hilbert space as

\begin{equation}
    \mathcal{H}_{TLS} = \text{span}\{|\short(t)\rangle,|\longg(t)\rangle\},
\end{equation}

spanned by the corresponding orthogonal states defined in Eq. (3). The Hilbert space defined in Eq.~(5) is a subspace of the electron's Hilbert space describing the laser-dressed continuum. It effectively captures the relevant manifold for HHG emission, as contributions from higher-order trajectories are exponentially suppressed \cite{Bengtsson2023}.

Building on \(\mathcal{H}_{TLS}\), we define the time-dependent electronic state as a dipole-weighted spinor that represents the coherent superposition of the short and long path contributions:

\begin{equation}
    |\psi_{TLS}(t)\rangle \equiv Re\lbrack{\hat{P}}_{s}\hat{z}|e\rangle + {\hat{P}}_{l}\hat{z}|e\rangle\rbrack = z_{\short}(t)|\short(t)\rangle + z_{\longg}(t)|\longg(t)\rangle,
\end{equation}

where \(z_{\short}(t)\) and \(z_{\longg}(t)\) are the time-dependent dipole responses (see Eq.~(S8) in the SM). While the spinor in Eq. (6) characterizes the TLS state transient sub-cycle evolution, the physical observables in HHG are generally frequency-resolved. We obtain the harmonic-order spinor of the APQ by applying a Fourier transform to the time-dependent state, Eq. (6), and sampling it at q\(\omega_{0}\) where \(\omega_{0}\) is the angular frequency of the pump beam and q is the harmonic order. That is, we define the frequency-resolved APQ spinor as

\begin{equation}
    \left| \psi_{APQ}\left( q\omega_{0} \right) \right\rangle = z_{\short}(q\omega_{0})|\short(q\omega_{0})\rangle + z_{\longg}(q\omega_{0})|\longg(q\omega_{0})\rangle,
\end{equation}

where \(z_{\short}(q\omega_{0})\) and \(z_{\longg}(q\omega_{0})\) are the Fourier transform of \(z_{\short}(t)\) and \(z_{\longg}(t)\), respectively. The APQ density matrix is thus given by

\begin{equation}
    \rho_{APQ}(q\omega_{0}) = |\psi_{APQ}(q\omega_{0})\rangle\langle\psi_{APQ}(q\omega_{0})| = {\hat{P}}_{s}\hat{z}\rho_{e}{\hat{z}}^{\dagger}{\hat{P}}_{s} + {\hat{P}}_{l}\hat{z}\rho_{e}{\hat{z}}^{\dagger}{\hat{P}}_{l} + {\hat{P}}_{s}\hat{z}\rho_{e}{\hat{z}}^{\dagger}{\hat{P}}_{l} + {\hat{P}}_{l}\hat{z}\rho_{e}{\hat{z}}^{\dagger}{\hat{P}}_{s}
\end{equation}

where \(\rho_{e} = |e\rangle\langle e|\) is the pure electron density matrix derived from the continuum state in Eq. (2). In this framework, the APQ represents the reduced state of the electron in the continuum as mapped onto the spectral dipole transition. Crucially, because the continuum wave packets span a broad energy range throughout the plateau, this formulation does not merely define a single isolated two-level system, but rather an ensemble of independent APQs encoded in the frequency domain, one for each harmonic order $q$. This formulation casts the recombination step as an intrinsic measurement process: the dipole operator effectively acts as a readout mechanism, projecting sub-cycle path-coherence onto the bound state and translating the electron's dynamical journey into observable radiation. Consequently, the high-harmonic spectrum serves as a qubit readout, where spectral yield modulations in the plateau provide a direct signature of the phase and coherence accumulated between the short and long trajectories in the laser-dressed continuum. Ultimately, the APQ framework establishes HHG as a formal platform for trajectory-based attosecond interferometry,
enabling the quantification of information loss through
frequency-resolved observables.

The dipole-weighted state in Eq. (6) represents the physical response of the system, which does not naturally preserve a unit norm under the no-depletion approximation. To formalize the APQ as a two-level system and map its dynamics onto the Bloch sphere, we define a normalized density matrix by explicitly dividing by the total dipole yield. This choice prioritizes the characterization of the relative path populations and phase coherence over the absolute harmonic intensity, providing a clear analytical route to investigate the trajectory-based subsystem. Consequently, the normalized frequency-domain APQ density matrix is defined by

\begin{equation}
    \rho_{APQ}\left( q\omega_{0} \right) = \begin{pmatrix}
\rho_{ss}\left( q\omega_{0} \right) & \rho_{sl}\left( q\omega_{0} \right) \\
\rho_{sl}^{*}\left( q\omega_{0} \right) & \rho_{ll}\left( q\omega_{0} \right)
\end{pmatrix},
\end{equation}

where the matrix elements are defined as \(\rho_{ss} = |z_{\short}(q\omega_{0})|^{2}/N(q\omega_{0})\), \(\rho_{ll} = |z_{\longg}(q\omega_{0})|^{2}/N(q\omega_{0})\), \(\rho_{sl} = z_{\short}^{*}(q\omega_{0})z_{\longg}(q\omega_{0})/N(q\omega_{0})\), and \(N(q\omega_{0}) = |z_{\short}(q\omega_{0})|^{2} + |z_{\longg}(q\omega_{0})|^{2}\) is a normalization factor. This framework establishes two complementary viewpoints: the frequency-domain APQ and the time-domain TLS. These dual descriptions facilitate different levels of physical insight. On one hand, the TLS acts as a probe of the underlying trajectory-coherent dynamics, capturing how the electron's journey in the continuum is mapped onto the time-dependent dipole response. On the other hand, the APQ provides the formal tools for an operational analysis, as it relates the coherent superposition of paths directly to the frequency-resolved observables of the HHG spectrum.

This reinterpretation of HHG as a trajectory-based interferometry establishes a bridge with the language of quantum optics and Quantum Information Science (QIS), as it allows us to explore HHG in terms of the quantum information encoded within its electronic paths. Within this framework, the interferometric visibility \emph{V} and path predictability \emph{P} emerge as fundamental metrics to characterize the wave-particle duality and the coherence of the electron's motion \cite{Englert1996,Drr1998,Qureshi2021}. Using the density matrix elements defined in Eq. (9), we express the interferometric visibility and predictability as:

\begin{equation}
    V(q\omega_{0}) = \frac{I_{\max} - I_{\min}}{I_{\max} + I_{\min}} = 2|\rho_{sl}(q\omega_{0})|,\quad P(q\omega_{0}) = |\rho_{ss}(q\omega_{0}) - \rho_{ll}(q\omega_{0})|.
\end{equation}
These two quantities, for a generic TLS, quantify the wave-like (\(V\)) and particle-like (\(P\)) behavior. Furthermore, they are complementary, satisfying \(V^{2} + P^{2} = 1\) for a pure state, where \(P\) characterizes how well one of the paths is known, and \(V\) measures the interference contrast in an interferometer \cite{Englert1996,Qureshi2021}. The qubit defined in Eq. (9) exhibits genuine interference between its two orthogonal states, with the short-long phase difference carrying a clear physical meaning that can be experimentally accessed through \(V\) and \(P\).

Building on these interferometric metrics, we introduce an operator basis that allows us to map the trajectory-based dynamics onto the Bloch sphere. It is important to note that this algebraic construction is representation-independent and can be applied to both the time-domain TLS and the frequency-domain APQ. While the formal definition of the Pauli operators remains the same, their physical interpretation adapts to the chosen domain: in the time domain, they track the transient sub-cycle evolution of the electron, whereas in the frequency domain, they characterize the properties of the emitted radiation.

By adopting the \(\{|\short(t)\rangle\),\(|\longg(t)\rangle\}\) basis as the natural framework for our qubit in either representation, the relevant observables are defined by the Pauli operators: \(\sigma_{z}\) measures the population imbalance (directly related to the predictability \emph{P}), while \(\sigma_{x}\) and \(\sigma_{y}\) quantify the real and imaginary parts of the path coherence, respectively (whose magnitude determines the visibility \emph{V}). This algebraic framework provides a compact way to track how variations in the driving laser properties (such as intensity, wavelength, or pulse shape) and decoherence channels modify the qubit state. The Pauli operators are defined as,

\begin{equation}
    \begin{aligned}
\sigma_{z} & = |\short(t)\rangle\langle \short(t)| - |\longg(t)\rangle\langle \longg(t)|, \\
\sigma_{y} & = i(|\longg(t)\rangle\langle \short(t)| - |\short(t)\rangle\langle \longg(t)|), \\
\sigma_{x} & = |\longg(t)\rangle\langle \short(t)| + |\short(t)\rangle\langle \longg(t)|.
\end{aligned}
\end{equation}

The respective Bloch vector components are given by \(r_{j} = Tr\lbrack\rho_{APQ}\sigma_{j}\rbrack\), with \(j = x,y,z\). The squared density matrix's trace, \(Tr\lbrack\rho^{2}\rbrack\), yields the qubit's purity which equals unity for a pure state and decreases as coherence is lost due to coupling to unobserved degrees of freedom or ensemble averaging.

The resulting state can therefore be represented by a Bloch vector within the unit sphere, whose direction encodes the relative populations and phase of the short--long superposition, while its length serves as a measure of the degree of coherence. Figures 1(b) and 1(c) illustrate this mapping for the APQ. Figure 1(b) shows the Bloch-sphere representation of a single trajectory superposition, with the poles identified as the limiting short and long path states. Figure 1(c) depicts ensemble averaging, where a distribution of Bloch vectors with different orientations yields a contracted mean vector, providing a geometric visualization of a mixed state with reduced coherence.

\begin{figure}[ht!]
    \centering
    \includegraphics[width=0.8\linewidth]{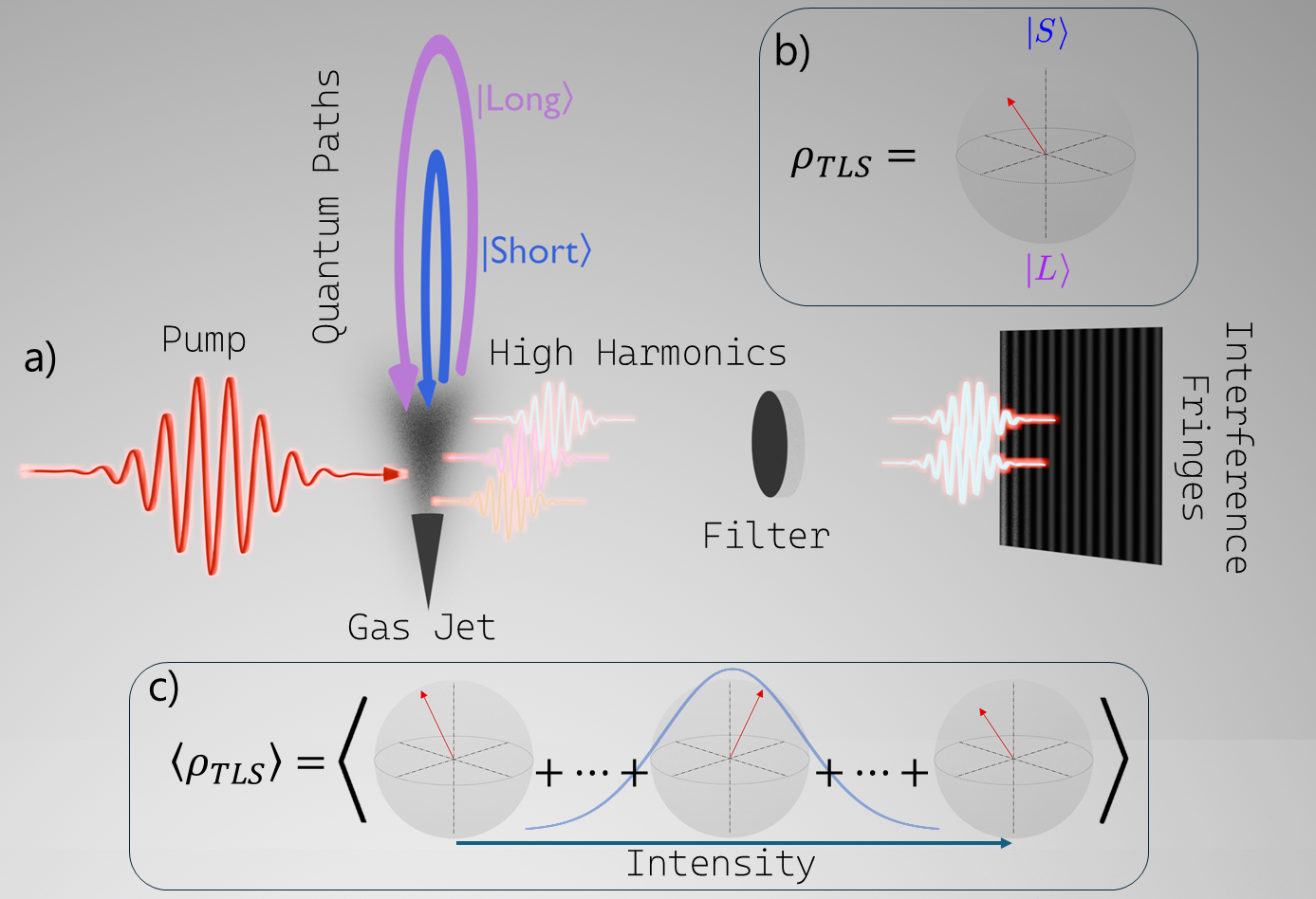}
    \caption{\textbf{Attosecond-path qubits in HHG.} (a) Schematic of a prototypical HHG experiment, together with the dominant quantum orbits that define the two-path states. (b) Bloch-sphere representation of the APQ density matrix (See Eq. (9)), and (c) of its ensemble average under shot-by-shot fluctuations. For each laser shot, a distinct effective two-level state is defined, and the ensemble average over these realizations yields the mixed state introduced in Eq. (26).}
    \label{fig:1}
\end{figure}

\section{APQ TOMOGRAPHY}

The APQ and its density matrix are directly measurable using established techniques in attosecond technology \cite{Biegert2021}. This experimental accessibility is fundamental to density matrix tomography (DMT), the protocol used to reconstruct the quantum state,\(\ \rho\), from a set of observables \cite{Nielsen2010,Monroe2002}. In the context of APQ, both the preparation and measurement of the short and long trajectories are accessible with current experimental tools \cite{Biegert2021}. Specifically, techniques such as phase matching, two-color (\(\omega - 2\omega\)) schemes, and intensity or phase scans enable controlled manipulation of the relative amplitude and phase between the short and long electron trajectories \cite{Brugnera2011,Zair2008,Shafir2012,Heyl2012,Schapper2010}, effectively allowing one to prepare different dynamical APQ states.

Eq. (9) define the APQ density matrix in the short--long basis. Here, the populations (\(\rho_{ss}\), \(\rho_{ll}\)) describe the relative contributions of each trajectory to the high-harmonic emission, and the coherence term (\(\rho_{sl}\)) encodes their ability to interfere. Importantly, these quantities are directly linked to experimentally measurable observables. The populations can be obtained from the relative signal associated with each trajectory, while the trajectory coherence is extracted from the interferometric visibility of the harmonic emission.

In the simplest case, neglecting all possible decoherence channels for the electron and given that the short and long trajectories dominate the harmonic yield, the harmonic spectrum \(H(\omega)\) arises from the coherent superposition of these two main contributions,

\begin{equation}
    H(\omega) = I_{\short} + I_{\longg} + 2\sqrt{I_{\short}I_{\longg}}cos(\phi),
\end{equation}

where \(I_{\short}\) and \(I_{\longg}\) are the individual intensities of the trajectories and \(\phi\) is their relative phase. This expression can be directly mapped onto the density matrix formalism as

\begin{equation}
    H(\omega) = \frac{\rho_{ss}(\omega) + \rho_{ll}(\omega) + 2\left| \rho_{sl}(\omega) \right|\cos(\phi)}{N},
\end{equation}

where \(\rho_{sl} = |\rho_{sl}|e^{i\phi}\) and $N$ is a normalization factor. Under ideal conditions, the coherence reaches its maximum value, \(|\rho_{sl}|\) =\(\sqrt{\rho_{ss}\rho_{ll}}\), corresponding to a pure state.

Reconstruction of \(\rho_{APQ}\) can be obtained by attosecond interferometry. The visibility \emph{V} yields \(|\rho_{sl}|\), the population imbalance \emph{P} yields \(\rho_{ss}{- \rho}_{ll}\). The remaining parameter, the coherence phase, \(\phi,\) is extracted from the phase shift of the interference term in Eq. (12) under a controlled scan of the relative trajectory phase. Thus, current technology should allow a complete reconstruction of the APQ state across the harmonic spectrum or as a function of external control parameters.

This tomographic approach also provides a direct way to identify and quantify decoherence. By comparing the reconstructed coherence \(\rho_{sl}\) with the maximum value expected for a pure state, \(\sqrt{\rho_{ss}\rho_{ll}}\), deviations can be interpreted as signatures of information loss due to coupling with additional degrees of freedom. Through systematic control over experimental parameters, DMT offers a practical and experimentally accessible route to probe decoherence in strong-field light--matter interactions.

\section{DEPHASING AND DECOHERENCE}

Standard models describing HHG, such as the semiclassical three-step model and the SFA, typically treat the emission process as a closed, single-electron system \cite{Lewenstein1994}. However, a more comprehensive physical picture requires recognizing that the emission process fundamentally behaves as an open quantum system. In this context, the trajectory-defined APQ acts as an open quantum subsystem that couples to environment-like degrees of freedom, including (but not limited to) the parent nucleus, neighboring electrons, and the radiation field, which may include unresolved soft photons \cite{Popruzhenko2018}. As the continuum electron becomes entangled with these variables, the reduced state of the APQ can no longer be represented as a pure state, but instead evolves into a mixed state. Within this framework, we focus on two distinct mechanisms of coherence loss: classical dephasing from ensemble averaging and genuine decoherence arising from the trace-out of unobserved environmental variables.

Dephasing emerges primarily from ensemble averaging over classical fluctuations that randomize the relative phase between the short and long trajectories. While the underlying dynamics of each individual realization remain formally unitary, the macroscopic interference signal is suppressed upon averaging \cite{Salires2001}. This channel includes contributions from spatial volume averaging across the gas jet and Coulomb-induced phase dispersion \cite{Gaarde2008,Popruzhenko2018}, as well as the energy-dependent phase dispersion caused by wavepacket spreading in the continuum \cite{Lewenstein1994,Salires2001}. Crucially, stochastic variations in the driving field intensity modify the accumulated dipole phase on a realization-by-realization basis, acting as a dominant source of classical noise in experimental settings.

In contrast, genuine decoherence arises when the electronic trajectories become entangled with unobserved degrees of freedom that carry which-path information \cite{Zurek2003}. Because the excursion of the electron trajectories occurs on a sub-cycle timescale, this interaction is far from the thermalization limit. The system does not reach a thermal equilibrium with its surroundings; instead, the coherence loss is governed by the rapid, non-equilibrium build-up of entanglement with internal ionic channels or the emitted radiation field \cite{Hernndez-Garca2013}. This process is intrinsically irreversible upon tracing over the environmental variables, even at the single-shot level \cite{Goh2015,Knzel2015}.

The distinction between these two channels carries significant operational implications for the characterization of attosecond states. While dephasing results from a lack of external parameter control, the resulting coherence loss can, in principle, be mitigated by conditioning the measurement on the fluctuating variable or by restricting the detection volume. In contrast, decoherence arising from the trace-out of intrinsic degrees of freedom represents a fundamental limit on the purity of the APQ that persists even in a single, perfectly controlled laser shot. In the following sections, we quantitatively investigate these two pathways of information loss by focusing on two representative case studies: the dephasing channel induced by shot-to-shot laser intensity fluctuations and the intrinsic decoherence channel arising from the trace-out of unresolved transverse electron momentum.

To formalize the distinction between these channels, we develop a general mathematical framework for quantifying the reduction in APQ purity, which underpins the numerical investigations that follow.

The dephasing channel is mathematically described through an ensemble average over a classically fluctuating parameter \(\lambda\), such as the laser intensity or the position of the emitters within the macroscopic generation volume. If \(P(\lambda)\) represents the probability distribution of this parameter and \(\rho(\omega,\lambda)\) is the pure-state density matrix for a specific realization, the observed state is given by the weighted integral:

\begin{equation}
    \rho_{avg}(\omega) = \int d\lambda\ P(\lambda)\rho(\omega,\lambda).\
\end{equation}

In this case, the loss of coherence in the off-diagonal terms of \(\rho_{avg}(\omega)\) arises because the different shots accumulate different relative phases, leading to destructive interference upon averaging even though each individual \(\rho(\omega,\lambda)\) remains
pure.

In contrast, the decoherence channel is formulated by treating the total system as a bipartite state \(\rho_{tot}\) that includes both the APQ degrees of freedom and an environmental degree of freedom, such as the transverse electron momentum \(\mathbf{p}_{\bot}\). The reduced density matrix of the APQ is obtained by performing a partial trace over the environment Hilbert space \(\mathcal{H}_{E}\):

\begin{equation}
    \rho_{APQ}(\omega) = {Tr}_{E}\left\lbrack \rho_{tot}(\omega) \right\rbrack = \int d\mathbf{p}_{\bot}\left\langle \mathbf{p}_{\bot}\left| \rho_{tot}(\omega) \right|\mathbf{p}_{\bot} \right\rangle.
\end{equation}

Unlike the dephasing channel, the reduction in purity here is a consequence of entanglement between the trajectory and the environment \cite{Nielsen2010}. This process ensures that even in a single laser shot, the APQ is in a mixed state if the environmental information is not resolved by the detector. Equations (14) and ( 15) establish the operational boundary of our study: the first allows us to model how external noise limits our control, while the second defines the fundamental limits imposed by the intrinsic multi-dimensional nature of the electron's motion.

\section{RESULTS}

In this section we investigate the attosecond path qubit dynamics. First, we establish the baseline pure-state dynamics of the APQ in the ideal single-atom, single-shot closed-system regime, where the electronic motion is mapped to a coherent orbit on the surface of the Bloch sphere. Second, we examine the dephasing channel induced by classical laser intensity fluctuations, demonstrating how fluctuating classical parameters reduces state purity and how this loss can be operationally reversed through post-selection. Finally, we investigate the decoherence channel arising from the trace-out of unresolved transverse electron momentum, which represents a fundamental and irreversible reduction in the purity of APQ. This progression allows us to quantitatively distinguish between experimental limitations and the intrinsic quantum constraints imposed by the intricate nature of HHG process.

The numerical results presented in the following sections are obtained using a SFA framework for a hydrogen-like atom driven by a mid-infrared linearly polarized laser field along z-direction

\begin{equation}
    E_{z}(t) = E_{0}f(t)cos(\omega_{0}t + \phi),
\end{equation}

where the peak amplitude of the laser electric field is \(E_{0} = 0.109\) a.u. (corresponding to a laser intensity of \(4.2 \times 10^{14}\) W/cm\(^{2}\)), and the laser frequency is \(\omega_{0} = 0.057\) a.u. (corresponding to a wavelength \(\lambda_{0} = 800\) nm). The pulse envelope \(f(t)\) has a trapezoidal shape:

\begin{equation*}
    f(t) = \left\{ \begin{matrix}
\sin^{2}\left( \frac{\omega_{0}t}{2n_{ramp}} \right) & 0 \leq t \leq \frac{2\pi n_{ramp}}{\omega_{0}}, \\
1 & \frac{2\pi n_{ramp}}{\omega_{0}} < t \leq \frac{2\pi\left( n_{cy} - n_{ramp} \right)}{\omega_{0}}, \\
\sin^{2}\left( \frac{\omega_{0}(T - t)}{2n_{ramp}} \right) & \frac{2\pi\left( n_{cy} - n_{ramp} \right)}{\omega_{0}} < t \leq T,
\end{matrix} \right.
\end{equation*}

where \(n_{ramp} = 1\), \(n_{cy} = 8\), and \(T = 2\pi n_{cy}/\omega_{0}\). The target is a hydrogen atom with \(I_{p} = 0.5\) a.u. For the dipole matrix element, we consider an \(s\)-like electron wavefunction for the electronic ground state

\begin{equation*}
    \psi(x) = \left( \frac{\alpha^{3/4}}{\pi^{1/2}} \right)e^{- \sqrt{\alpha}|x|},
\end{equation*}

where \(\alpha = 2I_{p}\). Since, within the SFA framework, continuum states are treated as plane waves, the dipole matrix element takes the analytical form \cite{Lewenstein1994}

\begin{equation*}
    \mathbf{d}(\mathbf{p}) = i\left( \frac{2^{7/2}\alpha^{5/4}}{\pi} \right)\frac{\mathbf{p}}{(p^{2} + \alpha)^{3}}.
\end{equation*}

The parameters given here set a classical cutoff around the 63rd harmonic \cite{Lewenstein1994}. Considering that the first harmonic fulfilling \(I_{p} > \omega\) is the 11th harmonic, and because the short-long basis is only defined in the plateau region, all the analysis shown in the following section is made for harmonics between the 13th and the 55th. Higher-order harmonics, according to the analysis and the Gabor maps shown in the SM, cannot be completely resolved into two different trajectories. Then, this selection assures a considerable distance from the classical cutoff and maintains the short-long basis definition.

\subsection{PURE STATE APQ DYNAMICS}

\begin{figure}[ht]
    \centering
    \includegraphics[width=0.8\linewidth]{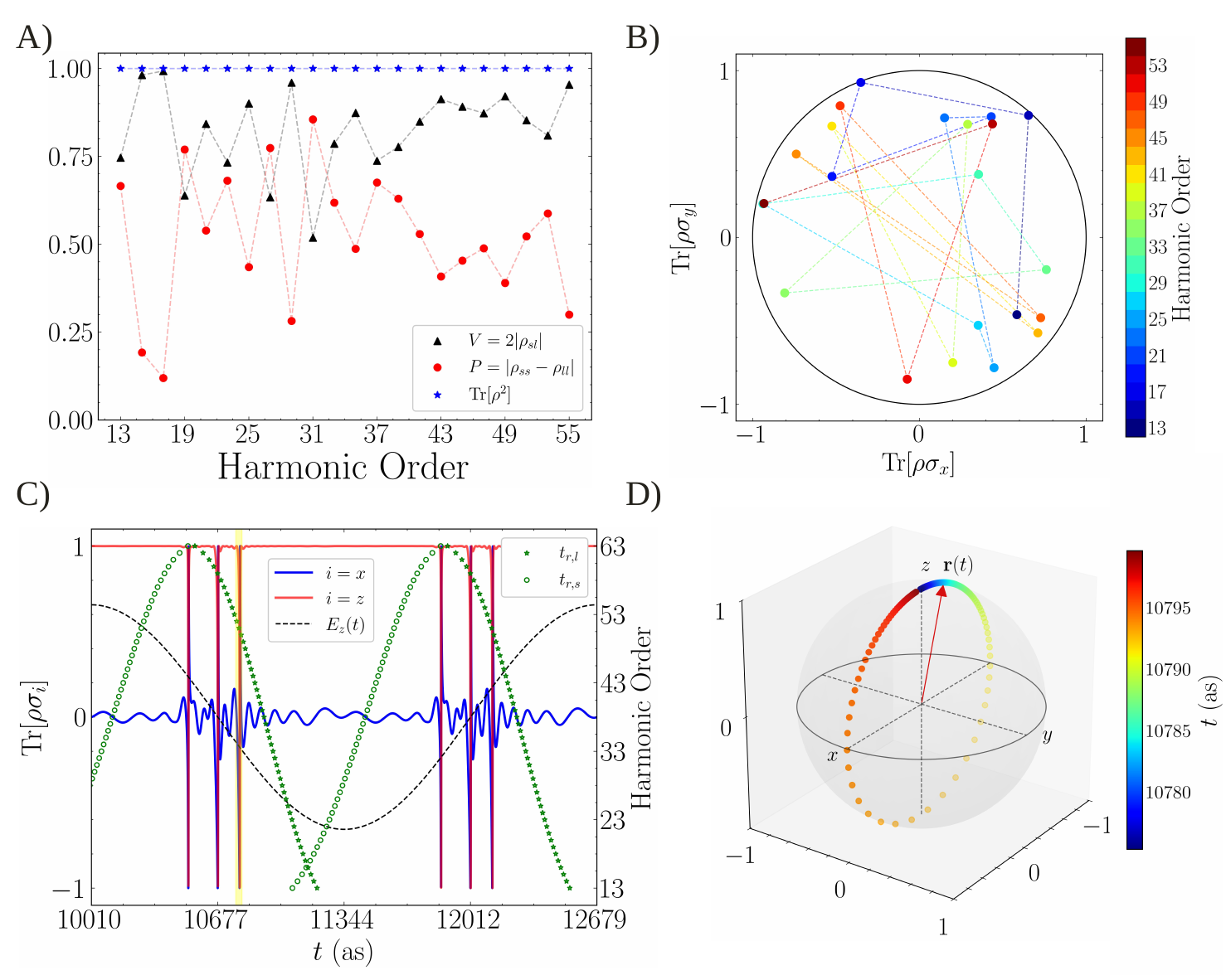}
    \caption{\textbf{APQ and TLS Bloch vector dynamics for pure-state density matrices}. (A) Interferometric visibility V (black triangles), predictability P (red circles), and state purity \(Tr\lbrack\rho^{2}\rbrack\) (blue stars) as functions of harmonic order for the APQ. (B) Complex coherence map showing the \(x\) and \(y\) Bloch-vector components as a function of harmonic order; the dashed lines trace the spectral evolution within the unit disk. (C) Real-time evolution of the Bloch-vector components during one optical cycle for the time-domain TLS, together with the semiclassical ionization (\(t_{i}\)) and recombination (\(t_{r}\)) times derived from saddle-point solutions (see SM). (D) Three-dimensional Bloch-sphere trajectory of the TLS during a \(\sim\)25 as window (highlighted in yellow in panel C), illustrating the rapid coherent excursion driven by the laser field defined in Eq. (16).}
    \label{fig:2}
\end{figure}

In the microscopic, single-atom limit, when all driving parameters are fixed and no environmental coupling is considered, the attosecond-path qubit remains in a pure state. Figure~\ref{fig:2} displays the corresponding Bloch vector dynamics for both the frequency-domain APQ and the time-domain TLS, illustrating the interferometric visibility,
predictability, and purity of the system.

Figure 2(a) demonstrates that the predictability \emph{P} (red markers) decreases toward zero as the harmonic order approaches the cutoff region. This trend reflects the coalescence of the short and long quantum orbits near cutoff, where their contributions become comparable and the path-population imbalance diminishes. Conversely, for harmonics within the plateau region, the predictability remains finite, reflecting a population imbalance dominated by the short trajectories, as shown in the SM.

The coherence of the system is analyzed in Figure 2(b), which maps the complex off-diagonal elements of the density matrix as a function of harmonic order through the \(x\) and \(y\) components of the Bloch vector. Both components exhibit rapid oscillations across the spectrum, a behavior that can be understood by expressing the dipole response in its complex form

\begin{equation}
    z_{j}(q\omega_{0}) = |z_{j}(q\omega_{0})|e^{i\phi_{j}(q\omega_{0})},
\end{equation}

where the phase \(\phi_{j}(q\omega_{0})\) can be expressed as \cite{Balcou1997}

\begin{equation}
    \phi_{j}(q\omega_{0}) = q\omega_{0}t_{r,j} - S(p_{s,j},t_{i,j},t_{r,j}),
\end{equation}

with \(j = \short\) (\(j = \longg\)) denotes the short (long) contribution, \(q = \omega/\omega_{0}\) the harmonic order, and \(t_{i}\) and \(t_{r}\) are the ionization and recombination and times obtained from the saddle-point solutions (see Eq. (S6) in the SM). Likewise, the \(x\) component of the Bloch vector can then be written as

\begin{equation}
    \text{Tr}\lbrack\rho_{APQ}\sigma_{x}\rbrack = 2\frac{|z_{\short}(q\omega_{0})||z_{\longg}(q\omega_{0})|}{|z_{\short}(q\omega_{0})|^{2} + |z_{\longg}(q\omega_{0})|^{2}}\cos(\phi_{\short}(q\omega_{0}) - \phi_{\longg}(q\omega_{0})),
\end{equation}

which naturally leads to an oscillatory dependence on the harmonic order \(q\). In an analogous way, the \(y\) component reads

\begin{equation}
    \text{Tr}[\rho_{APQ}\sigma_{y}] = - 2\frac{|z_{\short}(q\omega_{0})||z_{\longg}(q\omega_{0})|}{|z_{\short}(q\omega_{0})|^{2} + |z_{\longg}(q\omega_{0})|^{2}}\sin(\phi_{\short}(q\omega_{0}) - \phi_{\longg}(q\omega_{0})).
\end{equation}

The elements of the time-domain TLS density matrix are real-valued functions. Consequently, the Bloch vector becomes restricted to the\(x\)-\(z\) plane. Its \(x\) component takes the form

\begin{equation}
    \text{Tr}\lbrack\rho_{APQ}\sigma_{x}\rbrack = 2\frac{|z_{\short}(t)||z_{\longg}(t)|}{|z_{\short}(t)|^{2} + |z_{\longg}(t)|^{2}},
\end{equation}

which becomes significant only when the short and long trajectory contributions are of comparable magnitude.

Figure~\ref{fig:2}(c) shows the \(x\) and \(z\) components, together with the recombination times \(t_{r}\) for the short and long trajectories obtained from the SFA (see Eq.~(S6) in the SM). The dynamics are confined to the temporal region where the long trajectories recombine. Outside this interval, the \(z\) component remains close to unity (and the \(x\) component close to zero), indicating that the emission is dominated by the short trajectories. Only when the two contributions become comparable does the APQ exhibit a nontrivial evolution, which, as discussed above, takes place near the cutoff. Moreover, in Figure~\ref{fig:2}(d) we observe that during this interval the effective TLS rotates around the \(y\) axis, as expected for a unitary evolution on the Bloch sphere. This reveal pronounced sub-cycle dynamics in the time domain, with the Bloch vector completing a full rotation about the \(y\) axis within approximately one atomic unit of time, i.e., about \(24\) as (\(1\ as = 10^{- 18}\ s)\). The strong-field trajectory pair therefore behaves as an ultrafast TLS whose dynamics are encoded directly in the attosecond emission process.

\subsection{DEPHASING CHANNEL}

Following the general mathematical framework introduced in Eq. (14), we now investigate the dephasing channel induced by shot-to-shot laser intensity fluctuations. We model these fluctuations by an intensity Gaussian distribution with 1\% standard variation to Eq. (14)~\cite{Goh2015,Knzel2015}

\begin{equation}
    I \sim N\left( \mu = 4.2 \times 10^{14}\frac{W}{{cm}^{2}},\sigma = 0.01\mu \right),\ \ \ E_{0} = \sqrt{\frac{I}{I_{0}}},
\end{equation}

with \(I_{0} = 3.51 \times 10^{16}\ \frac{W}{{cm}^{2}}\) the atomic unit of intensity. The main effect of this fluctuation is to modify the dipole phase, while the dipole amplitude remains approximately unchanged \cite{Amini2019}. We therefore consider the amplitude \(|z(q\omega_{0})|\) intensity independent, and define the APQ state in the frequency domain for each laser shot as

\begin{equation}
    \rho_{APQ}(q\omega_{0},n) \equiv \frac{1}{N(q\omega_{0},n)}\begin{pmatrix}
|z_{\short}^{n}(q\omega_{0})|^{2} & z_{\short}^{n}(q\omega_{0})z_{\longg}^{n*}(q\omega_{0}) \\
z_{\short}^{n*}(q\omega_{0})z_{\longg}^{n}(q\omega_{0}) & |z_{\longg}^{n}(q\omega_{0})|^{2}\ 
\end{pmatrix},
\end{equation}

where \(n\) labels the shot number and \(N(q\omega_{0},n)\) is the normalization factor such that Tr\(\lbrack\rho(\omega,n)\rbrack = 1\). Utilizing Eq.~(17), the off-diagonal elements read

\begin{equation}
   [\rho_{APQ}(q\omega_{0},n)\rbrack_{sl} = \frac{1}{N(q\omega_{0})}|z_{\short}(q\omega_{0})||z_{\longg}(q\omega_{0})|e^{i(\phi_{\short}^{n}(q\omega_{0}) - \phi_{\longg}^{n}(q\omega_{0}))}.
\end{equation}

Because the trajectory-dependent dipole phases scale approximately linearly with laser intensity with nearly universal slopes \cite{Amini2019}, they directly inherit the statistical distribution of the driving field's fluctuations.

\begin{equation}
    \begin{aligned}
\phi_{short}(\omega) & = \alpha_{short}I \\
\phi_{long}(\omega) & = \alpha_{long}I,
\end{aligned}
\end{equation}

where \(\alpha_{short}\) and \(\alpha_{long}\) are trajectory-dependent
constants.

\begin{figure}[ht]
    \centering
    \includegraphics[width=0.8\linewidth]{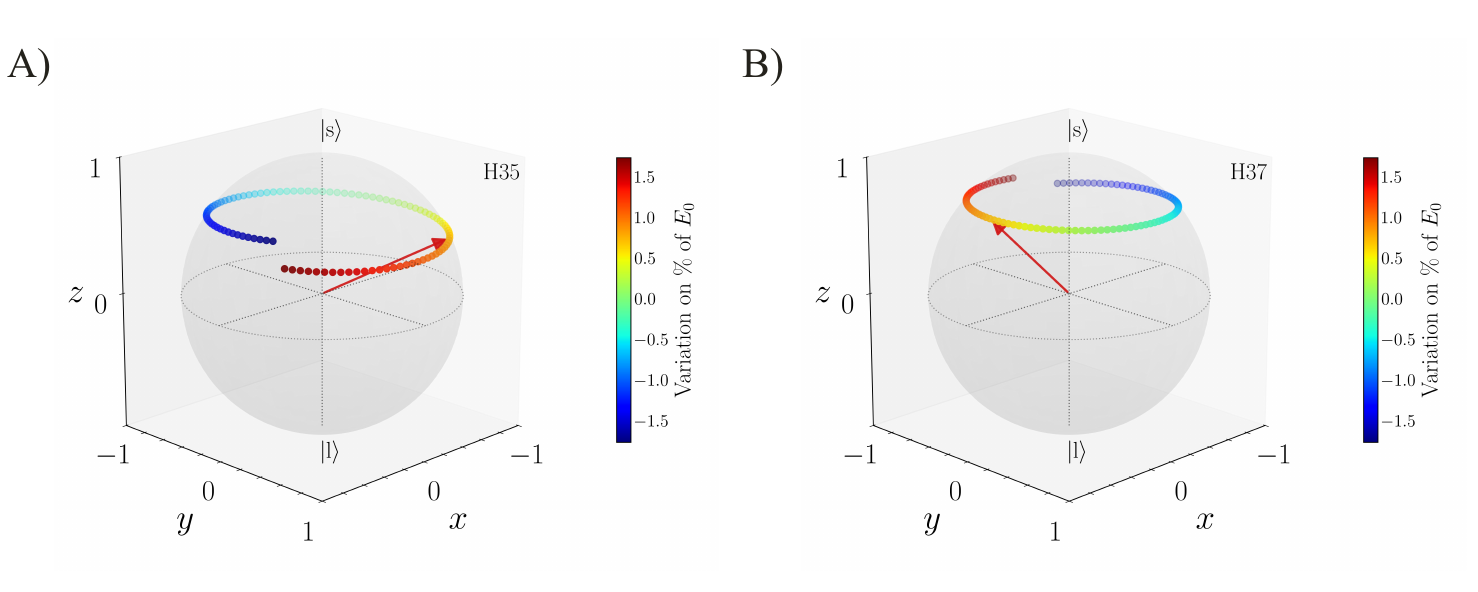}
    \caption{\textbf{APQ Bloch vector for different laser intensities.} (A) APQ Bloch vector for H35 and (B) H37, as a function of laser intensity. The scale is presented as a percentage of variation from the central value, \(E_{0} = 0.109\) a.u.}
    \label{fig:3}
\end{figure}

For any individual laser shot, a given peak intensity results in a definitive position of the Bloch vector with respect to the \(z\)-axis of the Bloch sphere. As the laser intensity varies from shot to shot, the Bloch vector corresponding to the density matrix in Eq.~(23) undergoes a rotation around the \(z\)-axis while the populations remain essentially unchanged, as shown in Figure~\ref{fig:3}.

The ensemble-averaged density matrix is obtained by averaging Eq. (23),

\begin{equation}
    \langle\rho_{APQ}\rangle \equiv \frac{1}{N}\begin{pmatrix}
|z_{\short}|^{2} & |z_{\short}||z_{\longg}|\langle e^{i(\phi_{\short}^{n} - \phi_{\longg}^{n})}\rangle \\
|z_{\short}||z_{\longg}|\langle e^{- i(\phi_{\short}^{n} - \phi_{\longg}^{n})}\rangle & |z_{\longg}|^{2} \\
 & 
\end{pmatrix},
\end{equation}

where \(\langle\ldots\rangle\) denotes the ensemble average as in Eq. (14). Using the intensity distribution in Eq.~(22), yields

\begin{equation}
    [\langle\rho_{APQ}(\omega)\rangle\rbrack_{sl} = \frac{1}{N(\omega)}|z_{\short}(\omega)||z_{\longg}(\omega)|e^{i\mu(\alpha_{\short} - \alpha_{\longg}) - \frac{1}{2}\sigma^{2}(\alpha_{\short} - \alpha_{\longg})^{2}}.
\end{equation}

Thus, the ensemble averaging acts as a dephasing channel (28). The populations remain unaffected, while the coherence between the two states is reduced. Physically, each laser shot introduces a random phase kick, corresponding to a random rotation around the \(z\)-axis of the Bloch sphere. When averaged over many shots, these random rotations lead to a progressive loss of phase information and thus a loss of coherence. Figure 4 illustrates the APQ and the TLS Bloch vector dynamics as a function of harmonic order and time, respectively. In Figure 4(a), we observe that the predictability \emph{P} remains essentially unchanged compared to the pure case in Figure 2(a). This directly reflects the nature of the dephasing channel, which suppresses off-diagonal terms while leaving the diagonal population elements intact. In contrast, the interferometric visibility \emph{V} is significantly reduced, capturing the loss of coherence induced by the averaging process. Similarly, comparing the complex coherence map in Figure 4(b) with the pure case in Figure 2(b), the ensemble-averaged state follows the same spectral trend, yet with systematically lower amplitudes.

Crucially, the most pronounced effect is observed in the state purity \(Tr\lbrack\rho^{2}\rbrack\). Across nearly the entire spectrum, the ensemble-averaged APQ resides within the Bloch sphere (\(V^{2} + P^{2} < 1\)), clearly indicating the transition to a mixed state and the loss of information inherent to the multi-shot averaging.

\begin{figure}[ht]
    \centering
    \includegraphics[width=0.8\linewidth]{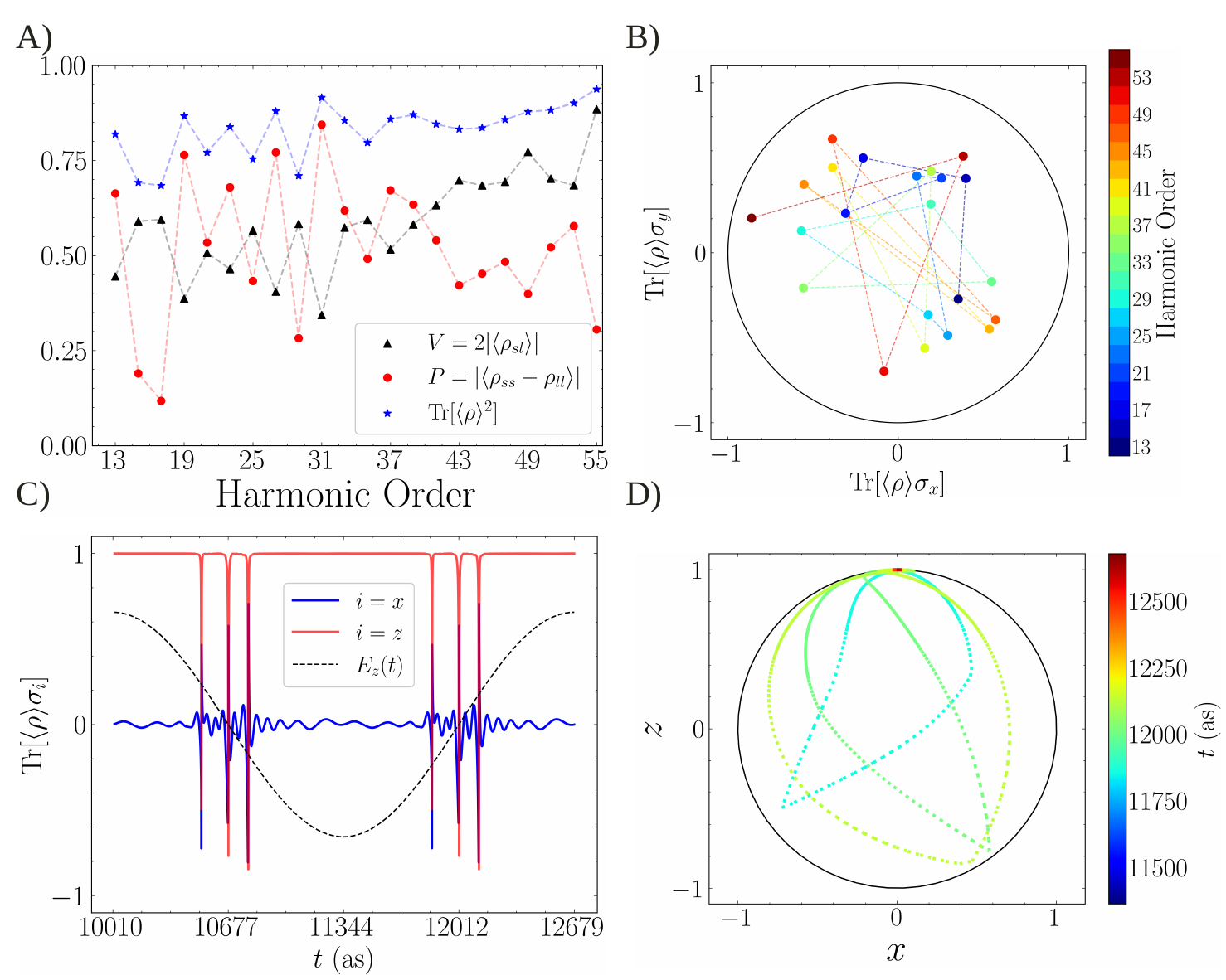}
    \caption{\textbf{APQ and TLS Bloch vector dynamics for the ensemble-averaged density matrix}. (A) Interferometric visibility V (black triangles), predictability P (red circles), and state purity \(Tr\lbrack\rho^{2}\rbrack\) (blue stars) as functions of harmonic order. (B) Complex coherence map showing the \(x\) and \(y\) Bloch-vector components as a function of harmonic order; Comparing with Figure 2(b), the contraction toward the center of the disk illustrates the dephasing-induced decoherence. (C) Real-time evolution of the ensemble-averaged Bloch-vector components during one optical cycle for the time-domain TLS. (D) Three-dimensional Bloch-sphere trajectory of the ensemble-averaged TLS over half an optical cycle, illustrating the stochastic reduction of the vector's magnitude within the Bloch sphere for the driving laser field defined in Eq. (16).}
    \label{fig:4}
\end{figure}

An analogous definition allows the mean density matrix in Eq.~(26) to be expressed in the time domain for the TLS. In Figure~\ref{fig:4}(c), we show the corresponding Bloch vector dynamics over one optical cycle. While the overall behavior resembles the pure case, an important difference emerges: the Bloch vector never reaches the point \((0,0, - 1)\) corresponding to the pure \(|\longg\rangle\) state. This feature becomes particularly clear in Figure~\ref{fig:4}(d), where the trajectory is shown for half optical cycle. Not only does the system fail to reach the \(|\longg\rangle\) state, but it also exhibits a clear loss of purity when entering the dynamical region. The accessible region of the Bloch sphere is therefore reduced due to the ensemble averaging. As in the pure case, the Bloch vector rotates around the \(y\)-axis, but now with a reduced length. The TLS still exhibits sub-cycle dynamics in the time domain, completing a full rotation on similar attosecond timescales, although with a diminished coherence.

As previously discussed, this reduction in purity underscores the operational nature of the dephasing channel. While Figure 4(b) shows a clear loss of coherence, it is crucial to remember that this is a consequence of the ensemble averaging over multiple laser shots. Each individual realization of the APQ remains intrinsically pure; however, the stochastic phase rotations around the \(z\)-axis, caused by intensity fluctuations, lead to a net reduction in the Bloch vector's length in the integrated signal. The information loss in this scenario is not fundamental: if the fluctuations were resolved on a shot-to-shot basis, the coherence shown in Figure 2 would be fully recovered, highlighting the distinction between this statistical dephasing and the intrinsic decoherence mechanisms discussed in the next section.

\subsection{DECOHERENCE CHANNEL}

Within the SFA, the \(z\)-component of the dipole amplitude before any saddle-point approximation is given by the Lewenstein integral \cite{Lewenstein1994} (See Eq.~(S1) in the SM). To model the effect of unresolved transverse degrees of freedom, we perform a partial saddle-point approximation only along the laser-polarization direction and keep the transverse-momentum integration explicit. This yields a transverse-momentum--resolved dipole response \(z(\omega,\mathbf{p}_{\bot})\) in which the semiclassical action separates naturally into parallel and transverse contributions,

\begin{equation}
    S\left( \mathbf{p},t,\tau \right) = S_{\parallel}\left( p_{z,s}(t,\tau),t,\tau \right) + S_{\perp}\left( \mathbf{p}_{\perp},t,\tau \right),
\end{equation}

with

\begin{equation}
    \begin{aligned}
S_{\parallel}(p_{z,s}(t,\tau),t,\tau) & = \int_{t - \tau}^{t}{dt^{'\frac{1}{2}}}(p_{z,s} + A_{z}(t'))^{2} + I_{p}\tau, \\
S_{\perp}(\mathbf{p}_{\perp},t,\tau) & = \frac{\tau}{2}\mathbf{p}_{\perp}^{2}.
\end{aligned}
\end{equation}

Given that \(\mathbf{p}_{\perp}\) is an independent variable (i.e., it does not vary with \(\tau)\), it takes the same numerical value for both the short and long trajectories. The phase difference between the two contributions therefore arises exclusively from the different excursion times \(\tau\). Treating \(\mathbf{p}_{\perp}\) as a free parameter thus separates the longitudinal action phase, governed by the excursion time, from transverse-spreading effects: different transverse momenta correspond to different emission angles and accumulate the additional phase, leading to angular dephasing when \(\mathbf{p}_{\perp}\) is unresolved. Notably, this transverse-spreading is absent in one-dimensional (1D) models, where the lack of transverse broadening artificially balances the weights of long and short trajectories (see SM for a detailed 1D analysis).

Defining a \(\mathbf{p}_{\perp}\)-resolved dipole amplitude allows the strong-field response to be viewed as a family of longitudinal quantum paths parametrized by \(\mathbf{p}_{\perp}\), enabling the role of transverse degrees of freedom to be examined before averaging. We then define the \(\mathbf{p}_{\perp}\)-resolved dipole response as

\begin{equation}
    \begin{matrix}
z(t,\mathbf{p}_{\perp}) & \equiv & i\int_{0}^{t}d\tau\sqrt{\frac{2\pi}{i\tau + \epsilon}}d_{z}^{*}(\mathbf{p}_{\perp},p_{z,s}(t,\tau) + \mathbf{A}(t))F_{z}(t - \tau) \\
 & & \times d_{z}(\mathbf{p}_{\perp},p_{z,s}(t,\tau) + \mathbf{A}(t - \tau))e^{- iS_{\parallel}(t,\tau) - i\frac{\tau}{2}\mathbf{p}_{\perp}^{2}} + c.c.
\end{matrix}
\end{equation}

The effective TLS and the APQ can now be constructed from the \(\mathbf{p}_{\perp}\)-resolved electron paths. Since the saddle-point condition is imposed only along the polarization axis, the stationary momentum \(p_{z,s}(t,\tau)\) specifies the longitudinal quantum orbits, namely the short and long trajectories, while \(\mathbf{p}_{\perp}\) remains a conserved canonical momentum throughout the continuum propagation. In this framework, the continuum electron state separates into longitudinal and transverse components. Accordingly, the total state is expressed as a coherent superposition over all transverse momentum configurations

\begin{equation}
    \begin{matrix}
|e\rangle = e^{iI_{p}t}|\phi_{0}\rangle + \int d^{2}\mathbf{v}_{\perp}\int dv_{z}b(\mathbf{v},t)|v_{z}\rangle|\mathbf{v}_{\perp}\rangle,
\end{matrix}
\end{equation}

where the integration is now split along the direction that encodes the longitudinal quantum orbits, represented by \(v_{\parallel}\), and the transverse direction represented by \(\mathbf{v}_{\perp}\). In a typical experimental configuration, the spectrometer axis is aligned with the laser polarization direction. Then, the detector collects radiation within a finite angular acceptance cone, which amounts to integrating over the corresponding distribution of transverse momenta. As the transverse emission angles (that correspond to orthogonal far-field spatial modes) are not coherently recombined by the detection optics, this angular integration is equivalent to a partial trace over the \(\mathbf{p}_{\bot}\)~degrees of freedom. The transverse momentum thus acts as an unresolved environment whose influence on the measured signal is averaged out, leaving only the longitudinal quantum-path information accessible. As before, we redefine the short-long basis as follows

\begin{equation}
    \begin{matrix}
|\short(t,\mathbf{p}_{\perp})\rangle & \equiv & \int_{\Omega_{s}}^{}{dp_{z}}b(\mathbf{p} + \mathbf{A}(t),t)\left| p_{\parallel}\mathbf{+}A(t) \right\rangle\exp\left( - iS_{\mathbf{p}}(t) \right), \\
|\longg(t,\mathbf{p}_{\perp})\rangle & \equiv & \int_{\Omega_{l}}^{}{dp_{z}}b\left( \mathbf{p} + \mathbf{A}(t),t \right)\left| p_{\parallel}\mathbf{+}A(t) \right\rangle\exp\left( - iS_{\mathbf{p}}(t) \right).
\end{matrix}
\end{equation}

Since \(\mathbf{p}_{\perp}\) acts as a continuous label for the longitudinal quantum trajectories, the APQ state is obtained by tracing over the transverse momentum degree of freedom,

\begin{equation}
    \begin{aligned}
\rho_{APQ}(\omega) & \equiv \frac{1}{N(\omega)}\int d^{2}p_{\perp}\begin{pmatrix}
|z_{\short}(\omega,p_{\perp})|^{2} & & z_{\short}(\omega,p_{\perp})z_{\longg}^{*}(\omega,p_{\perp}) \\
z_{\short}^{*}(\omega,p_{\perp})z_{\longg}(\omega,p_{\perp}) & & |z_{\longg}(\omega,p_{\perp})|^{2}
\end{pmatrix}
\end{aligned}
\end{equation}

with \(N(\omega) = \int d^{2}p_{\perp}(|z_{\short}(\omega,p_{\perp})|^{2} + |z_{\longg}(\omega,p_{\perp})|^{2})\) a normalization factor. Here, the integration is carried out element by element over the density matrix. This integration over the transverse momentum arises from the trace-out operation, effectively discarding the information associated with this degree of freedom (see SM for a detailed derivation). Therefore, the loss of such information behaves as a decoherence channel for the \(\mathbf{p}_{\perp}\)-resolved APQ, leading to a reduction of both the purity and the coherence of the system. From an operational perspective, one must therefore specify integration boundaries for the transverse momentum. Although the formal integration extends over all real values, in practice only a finite region around \(|\mathbf{p}_{\perp}| = 0\) contributes significantly. Indeed, the \(\mathbf{p}_{\perp}\)-resolved dipole amplitude in Eq. (30) is a decreasing, although oscillatory, function of \(|\mathbf{p}_{\perp}|\), reflecting the fact that the dipole matrix element itself decreases with increasing \(|\mathbf{p}_{\perp}|\). For \(|\mathbf{p}_{\perp}| > 1\) a.u., the dipole amplitude is already suppressed by roughly three orders of magnitude, so contributions from larger transverse momenta can be neglected.

Figure 5 displays the calculated (Eq. (33)) APQ and TLS Bloch vector dynamics as a function of harmonic order and time, respectively. In Figure 5(a), the predictability \emph{P} is both shifted and rescaled compared to the pure case, reflecting how the \(\mathbf{p}_{\perp}\)\hspace{0pt}-integration reweights the short and long contributions through the integrated intensities\(\ \int d^{2}p_{\bot}{\mid z_{s,l}(\omega,p_{\bot}) \mid}^{2}\). In particular, for most harmonic orders the balance shifts toward the long trajectory. Figure \ref{fig:6} clarifies this behavior by showing the \(\mathbf{p}_{\mathbf{\bot}}\)\hspace{0pt}-resolved dipole for two representative harmonics. Although long trajectories experience more rapid phase oscillations due to their larger excursion times, they retain a higher relative weight in the integrated signal. Therefore, when integrating over \(\mathbf{p}_{\bot}\), the long trajectories experience enhanced destructive interference due to the rapid oscillations of the phase term \(e^{- i\tau p_{\bot}^{2}/2}\), and are therefore strongly suppressed in the total signal. In contrast, when the transverse momentum is resolved, this cancellation is avoided, allowing the long contribution to retain a larger relative weight and become dominant.

The interferometric visibility \emph{V} and the state purity \(Tr\lbrack\rho^{2}\rbrack\) clearly reflect this decoherence channel, exhibiting a pronounced reduction across the entire spectrum. Notably, neither \emph{V} nor \emph{P} reaches unity, even for the lower-order harmonics. Furthermore, while the complex coherence map in Figure 5(b) follows the same spectral trend as the pure case, the decoherence channel induces a systematic reduction in the oscillation amplitudes. The corresponding Bloch vector therefore contracts toward the center of the sphere, indicating a substantial and permanent loss of quantum coherence due to the integration over the transverse momentum degrees of freedom.

\begin{figure}[ht]
    \centering
    \includegraphics[width=0.8\linewidth]{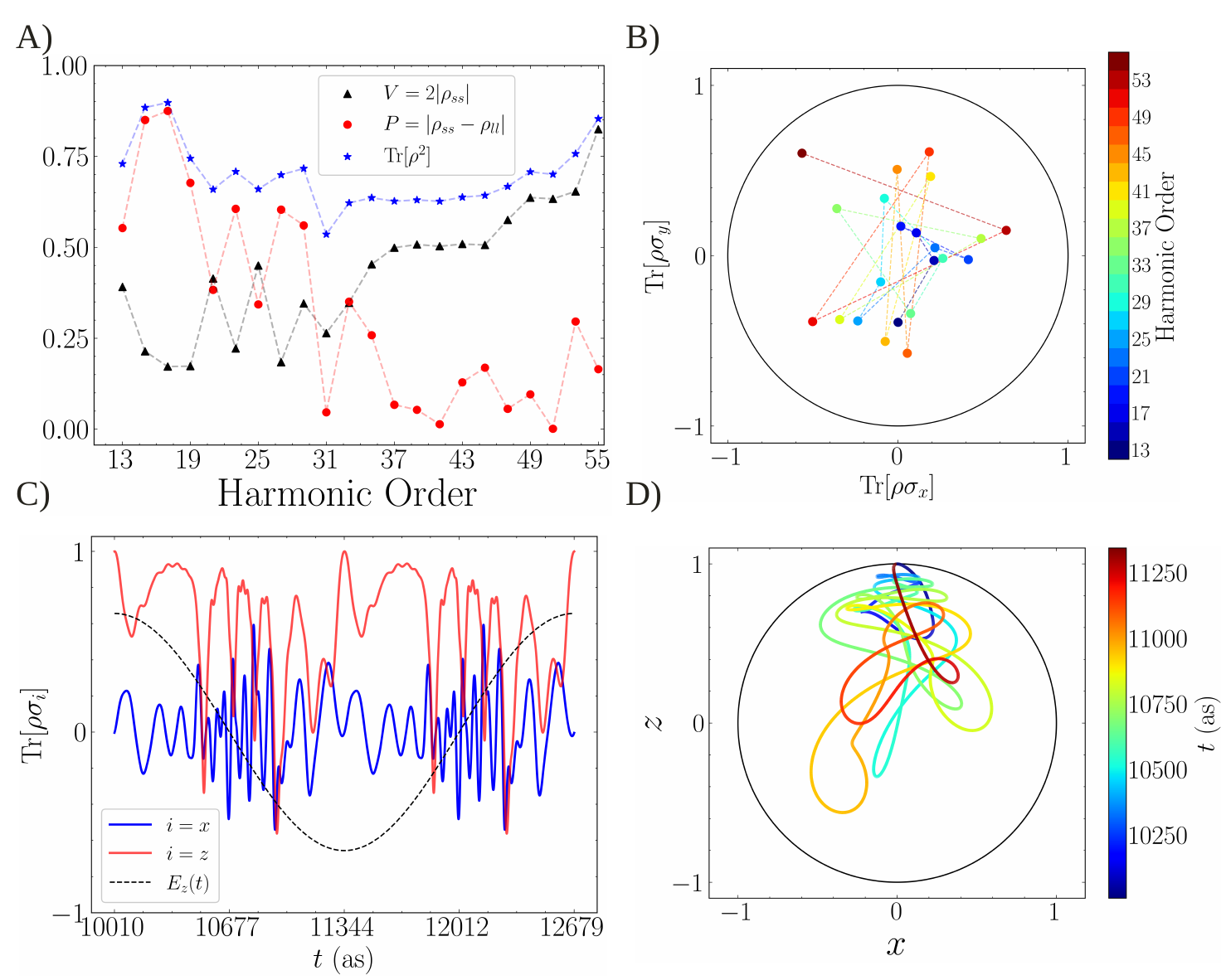}
    \caption{\textbf{Bloch vector dynamics after tracing over transverse momentum}. (A) Interferometric visibility V (black triangles), predictability P (red circles), and state purity \(Tr\lbrack\rho^{2}\rbrack\) (blue stars) as functions of harmonic order for the integrated APQ. (B) Complex coherence map representing the x and y Bloch-vector components as a function of harmonic order; Comparing with Figs. 2(b) and 4(b), the reduction in oscillation amplitudes illustrates the intrinsic decoherence induced by the trace-out of transverse degrees of freedom. (C) Real-time evolution of the integrated Bloch-vector components over one optical cycle for the time-domain TLS. (D) Three-dimensional Bloch-sphere trajectory of the TLS over half an optical cycle, illustrating the decoherence-driven path within the Bloch sphere for the driving laser field defined in Eq. (16).}
    \label{fig:5}
\end{figure}

In Figure~\ref{fig:5}(c), we show the corresponding TLS Bloch vector trajectory over one optical cycle. The \(x\) and \(z\) components remain half-cycle periodic, but the trajectory now exhibit a completely different landscape and a richer dynamic. The most significant change relative to the pure case is that the system never reaches the \(|long\rangle\) state; instead, it evolves continuously throughout the entire cycle. In the pure case, the system experiences a stationary behavior during a large portion of the cycle, remaining in the \(|\short\rangle\) state. By contrast, here the system reaches the \(|\short\rangle\) state once per half-cycle. This is shown in Figure~\ref{fig:5}(d) which zooms in on the dynamics over a half-cycle. Compared to the pure and dephasing cases, it is evident that the TLS undergoes a significant loss of purity within the decoherence channel. Accordingly, the evolution is no longer a simple rotation about the y axis with a fixed sense; instead, completing a full dynamical loop now requires an entire half-cycle.

\begin{figure}[ht]
    \centering
    \includegraphics[width=0.8\linewidth]{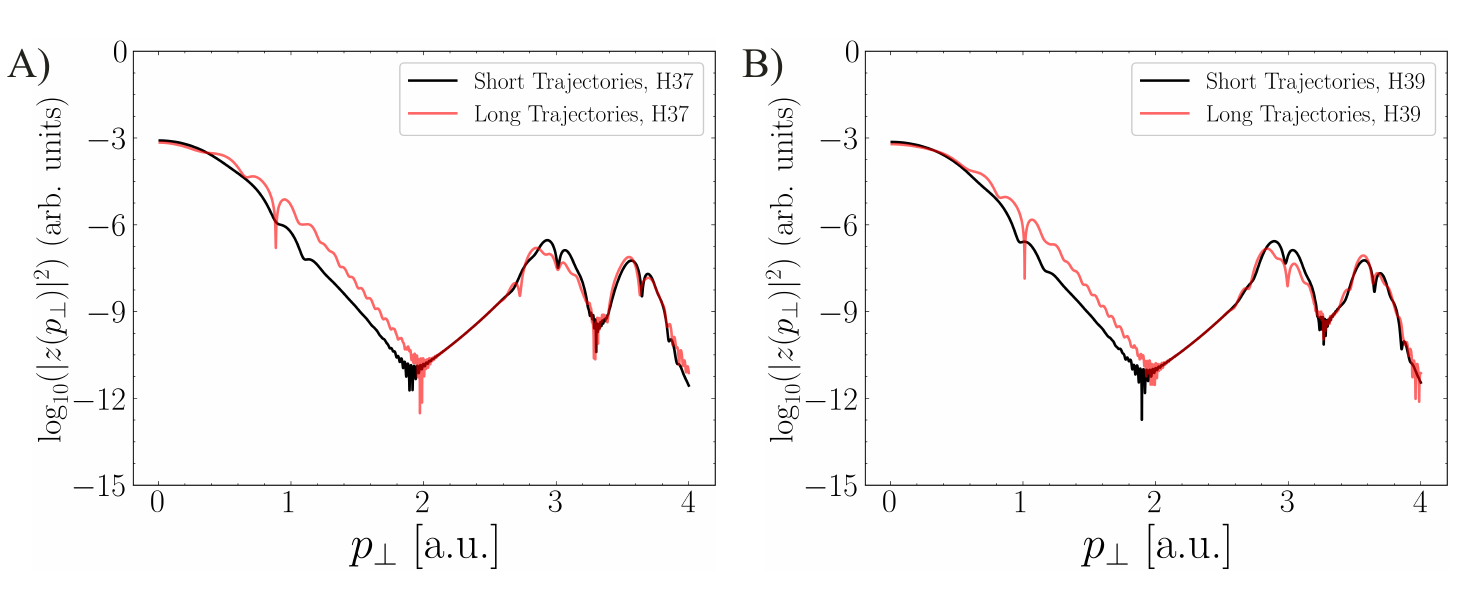}
    \caption{ \(\mathbf{p}_{\mathbf{\bot}}\)\textbf{-dependent dipole moment as a function of} \(\mathbf{p}_{\mathbf{\bot}}\) \textbf{for short and long trajectories}. (A) For H37 and (B) H39.}
    \label{fig:6}
\end{figure}

\section{CONCLUSIONS AND OUTLOOK}

In this work, we have introduced a trajectory-based description of HHG in which the dominant short and long electron paths define an effective two-level subsystem. Because these contributions can be experimentally distinguished, they naturally span a measurement-defined Hilbert space, allowing HHG emission to be described in terms of a trajectory-resolved density matrix. Within this framework, the relative amplitudes and phases accumulated along the two trajectories define an attosecond-path qubit (APQ), establishing a direct mapping between semiclassical electron dynamics and effective two-level quantum behavior. This formulation enables the use of quantum-information tools to characterize strong-field dynamics in a compact and operational way, expressing populations, coherences, and information loss in terms of Bloch vectors, Pauli operators, and purity measures. APQ framework provides a rigorous method for distinguishing between different mechanisms of coherence loss. We have shown that shot-to-shot fluctuations of the driving field intensity act as a classical dephasing channel that suppresses the off-diagonal elements of the density matrix. This type of coherence loss can be operationally reversible through conditioning the measurement on the fluctuating parameter, effectively restoring the purity of the state through post-selection. In contrast, the trace-out of unresolved degrees of freedom, such as the transverse momentum of the electron, represents a genuine decoherence channel. This process arises from the intrinsic entanglement between the electronic pathway and its environment, imposing a fundamental limit on the purity of the APQ that persists even in a single, perfectly controlled laser shot.

Importantly, the state of the APQ can be reconstructed from standard HHG measurements forming a protocol for APQ tomography. By utilizing attosecond technology \cite{Biegert2021} and established measurement techniques, both population imbalances and interferometric visibility are experimentally accessible. This provides a direct route to quantify the transition from pure to mixed states in strong-field electron dynamics, bridging the gap between attosecond science and quantum information theory.

The establishment of the APQ as a controllable two-level subsystem provides a concrete operational building block for the recently envisioned field of Attosecond Quantum Information Science (ATTOQUIS) \cite{Stammer2023}. By shifting the description of strong-field processes from semiclassical trajectories to a rigorous Hilbert-space representation, we move toward a regime where electronic motion can be treated as a controllable quantum resource. This framework allows for the translation of attosecond spectroscopy into a language of information flow, where the purity and coherence of the electronic state become primary observables. We discuss below several directions where this trajectory-based density matrix description can be extended to scale toward multi-qubit architectures and explore fundamental aspects of quantum mechanics.

While we treated here a two-level subsystem, corresponding to the HHG emission, the underlying dynamics are more accurately described by an evolving electronic qutrit spanned by the basis \(\{|\phi_{0}\rangle,\ |\short\rangle,\ |\longg\rangle\}\), incorporating the ground state as a dynamic participant in the coherence evolution. During the sub-cycle journey, the electronic state exists as a coherent superposition across all three channels, where \(|\phi_{0}\rangle\) provides the phase reference for ionization and the terminal point for recombination. This qutrit representation allows for the quantification of Stark effect and ground-state depletion and the tracking of population flux between the bound and continuum states. By treating the system as a qutrit, one should be able to resolve the "continuum journey" in its entirety, providing a complete account of the electronic state's purity before the projective "readout" into HHG radiation.

Our framework distinguishes between the electronicTLS, the coherent matter-state formed during the electron's excursion, and the photonic APQ, the state of the emitted radiation. The electronic TLS carries the direct "memory" of the continuum journey, including its entanglement with many-body environments. The recombination process then acts as a mapping operator which translates the electronic coherence into the optical properties of the HHG radiation. This distinction is vital for understanding information transduction: while the electronic TLS may suffer from intrinsic decoherence (the "decoherence scar"), the photonic APQ is the observable signature that allows for the operational reconstruction of that state. Can we design protocols to transfer complex electronic coherence structures to the HHG radiation?

The reduction in APQ purity, the "decoherence scar", serves as a high-fidelity probe of the quantum environment at the temporal frontier. Because the trace-out decoherence channel is sensitive to any degree of freedom that entangles with the electron, measuring the purity of the APQ provides a "coherence-contrast" spectroscopy of many-body correlations. This includes the influence of multi-electron screening, the coupling to nuclear degrees of freedom, and the fundamental limit imposed by the emission of QED soft photons through infrared dressing. We expect each of these interactions to leave a distinct imprint on the qubit's state, effectively turning the APQ into a microscopic sensor. This metrology can be further enriched by employing quantum light drivers, such as squeezed vacuum. The resulting decoherence scars would encode the sub-cycle entanglement between the electronic pathways and the quantized driving field.

The APQ framework may be suited for fundamental tests of quantum mechanics and investigation of quantum thermodynamics on the currently shortest accessible timescales \cite{Kosloff2013}. The rapid, action-driven phase evolution of the APQ, driven by extremely large instantaneous energy splitting, may allow exploration of Quantum Speed Limits (QSLs) in strongly non-stationary systems \cite{Deffner2017,Wang2026}. By characterizing the rate of state evolution against the accumulation of semiclassical action, one can probe the ultimate physical bounds of quantum evolution speed. Parallel to these limits, the APQ provides a testbed for studying the entropy production associated with decoherence. Measuring the pure-to-mixed state transition allows for an assessment of the "energetic price" of information loss in ultrafast strongly driven systems without thermalization. Furthermore, the accessibility of trajectory-resolved phases suggests a route toward testing temporal non-classicality via Leggett-Garg Inequalities (LGIs), testing the limits of macrorealism at the atomic scale \cite{Leggett1985,Emary2014}.

The trajectory-resolved framework is inherently scalable toward higher-dimensional and multi-qubit architectures. By employing multicolor driving fields or non-collinear geometries, the number of coherent electronic pathways can be increased to generate attosecond qudits, allowing for complex state encoding within a single laser cycle. The high-harmonic frequency comb naturally forms a one-dimensional quantum network, where each harmonic order provides a discrete readout of the underlying electronic coherence. Future research could explore the generation of entangled cluster states across this comb, utilizing macroscopic propagation to map microscopic electronic entanglement onto structured macroscopic optical modes. Such an architecture may allow HHG to transfer sub-cycle electronic dynamics into scalable, multi-dimensional photonic states.

Attosecond qubits offer a uniquely fast and controllable setting for exploring and tailoring coherence, dephasing, and decoherence at the fundamental timescale of electronic motion. Thus, we expect new directions for probing quantum dynamics in extreme, ultrafast, and strongly driven regimes that are otherwise inaccessible in other quantum systems.

\bibliography{refs}
\section*{Acknowledgments}

\subsection*{Funding:}

Quantum Science and Technology-National Science and Technology Major Project~(Grant No. 2025ZD0301000)
Guangdong Provincial Quantum Science Strategic Initiative (Grant No. GDZX2504001)
National Key Research and Development Program of China grant 2023YFA1407100 (MFC), 
Guangdong Province Science and Technology Major Project, Future functional materials under extreme conditions, grant 2021B0301030005 (MFC),
National Natural Science Foundation of China grant 12574092 (MFC).

\subsection*{Author contributions:}

O.C devised the concept of attosecond path qubits \cite{Cohen2026}. A.M. and C.B. developed the theoretical framework. A.M. performed the calculations and wrote the first version of the manuscript. O.C. and M.F. C. supervised the project. All authors discussed the results and contributed to the final version of the manuscript.

\subsection*{Competing interests:} All other authors declare they have no competing interests.

\subsection*{Data and materials availability:} All data and code needed to evaluate and reproduce the results in the paper are present in the paper and/or the SM.

\end{document}

\maketitle
\beginsupplement

\begin{center}
    \small
    \textbf{Corresponding authors:} \\
    Email: \href{mailto:marchisio.andres@gtiit.edu.cn}{marchisio.andres@gtiit.edu.cn} \\
    Email: \href{mailto:oren@technion.ac.il}{oren@technion.ac.il}
\end{center}

\vspace{2em} % Espacio antes de que empiece el contenido del suplemento

\begin{abstract}
    In this supplementary material, we provide extended derivations, numerical approach, and complementary analyses to support the results reported in the Main Text. We begin with a description of the theoretical and numerical frameworks used to evaluate the strong-field dipole responses corresponding to short and long electronic trajectories, establishing the formal basis states for the Attosecond-Path Qubit formulation. To expand our study of laser-intensity fluctuations and the resulting dephasing mechanisms, we show the APQ Bloch sphere dynamics for multiple harmonic orders. We then assess the physical origins of decoherence by introducing a comparative one-dimensional simulation, which explicitly highlight how integrating over the transverse momentum degrees of freedom acts as an intrinsic trace-out channel. Finally, we provide a step-by-step mathematical derivation of the reduced, partially-traced APQ density matrix formulated in Eq. (33) of the Main Text.
\end{abstract}

\section{Derivation of Coefficients of Eq. (6) and (7) of the Main Text}

The structural path decomposition of the strong-field dipole response
provides the theoretical foundation for both the two-level system (TLS)
description and the Attosecond-Path Qubit (APQ) formulation. Within the
Single-Active-Electron (SAE) and Strong-Field Approximations (SFA), the
time-dependent dipole response driven by a laser field linearly
polarized along the \(z\)-axis is given by the Lewenstein model
\cite{Lewenstein1994}

\begin{equation}
    \begin{matrix}
z(t) & = & i\int_{0}^{t}d\tau\int_{}^{}{d^{3}\mathbf{p}}d_{z}^{*}(\mathbf{p} + \mathbf{A}(t))E_{z}(t - \tau)d_{z}(\mathbf{p} + \mathbf{A}(t - \tau)) \\
 & & e^{- iS(\mathbf{p},t,\tau)} + c.c.,
\end{matrix}
\end{equation}
where \(\tau\) is the excursion time, \(\mathbf{p}\) is the canonical momentum, \(\mathbf{d}(\mathbf{v}) = \langle\mathbf{v}|z|\phi_{0}\rangle\) is the dipole matrix element for the bound-free transition, and \(\mathbf{A}(t)\) is the laser vector potential. In Eq.~(S1), the semiclassical action, \(S(\mathbf{p},t,\tau)\), is given by

\begin{equation}
    S(\mathbf{p},t,\tau) = \int_{t - \tau}^{t}\frac{1}{2}(\mathbf{p} + \mathbf{A}\left( t' \right))^{2}dt' + I_{p}\tau,
\end{equation}
with \(I_{p}\) the ionization potential. The integral in Eq. (S1) has a straightforward interpretation: \(E_{z}(t - \tau)d_{z}(\mathbf{p} + \mathbf{A}(t - \tau))\) gives the probability amplitude for a bound electron to make the transition to the continuum at the ionization time \(t - \tau\), with canonical momentum \(\mathbf{p}\). The electron is then propagated until time \(t\), where it acquires a phase factor equal to the semiclassical action \(S(\mathbf{p},t,\tau)\). Within the SFA, the effects of the atomic potential are neglected during the electron trip in the continuum, so that \(S(\mathbf{p},t,\tau)\) describes the motion of a free electron moving solely in the laser field, with a constant canonical momentum \(\mathbf{p}\). Finally, the electron recombines at time \(t\) with a probability amplitude equal to
\(d_{z}^{*}(\mathbf{p} + \mathbf{A}(t))\). In this sense, Eq.~(S1) provides the semiclassical formulation of the well-known three-step picture of HHG \cite{Kulander1993,Corkum1993} .

Due to the highly oscillatory nature of the integrand in Eq.~(S1), the main contributions to the dipole response arise from the stationary points of the action with respect to both \(\mathbf{p}\) and \(\tau\). Therefore, it can be solved by the saddle-point approximation \cite{Lewenstein1994}. Applying the stationary-phase condition exclusively to the spatial momentum coordinate \(\mathbf{p}\), the dipole response yields

\begin{equation}
    \begin{matrix}
z(t) & = & i\int_{0}^{t}d\tau\left( \frac{2\pi}{i\tau + \epsilon} \right)^{3/2}d_{z}^{*}(\mathbf{p}_{s}(t,\tau) + \mathbf{A}(t))E_{z}(t - \tau) \\
 & & d_{z}(\mathbf{p}_{s}(t,\tau) + \mathbf{A}(t - \tau))e^{- iS(t,\tau)} + c.c.,
\end{matrix}
\end{equation}
where \(\mathbf{p}_{s}\) denotes the saddle-point canonical momentum determined from the stationary-phase condition of the rapidly oscillating action with respect to \(\mathbf{p}\)

\begin{equation}
    p_{z} = - \frac{1}{\tau}\int_{t - \tau}^{t}A_{z}(t')dt',\quad p_{x} = p_{y} = 0.
\end{equation}

As a direct consequence of applying this saddle-point condition, the integration over the transverse momentum components (\(p_{x}\) and \(p_{y}\) directions) is analytically evaluated, appearing explicitly as an effective wavefunction broadening given by the prefactor
\(\left( \frac{2\pi}{i\tau + \epsilon} \right)^{3/2}\) in Eq.~(S3). Here, \(\epsilon\) is a positive infinitesimal constant arising from the Gaussian integration over the momentum components around the saddle point. The emitted radiation intensity is given by the celebrated Larmor formula, being proportional to the Fourier transform of Eq.~(S3), \(I \propto |\omega^{2}z(\omega)|^{2}\), where \(z(\omega)\) is given by

\begin{equation}
    \begin{matrix}
z(\omega) & = & \frac{1}{\sqrt{2\pi}}\int_{- \infty}^{\infty}z(t)e^{i\omega t} + c.c.
\end{matrix}
\end{equation}

Further approximations can be made, as noticing that the phase in Eq.~(S5), \(S(t,\tau) - \omega t\), is still a highly-oscillatory function of \(t\) and \(\tau\) \cite{Lewenstein1994}. The corresponding stationary-phase
conditions with respect to the remaining temporal variables yield the saddle-point energy conservation equations

\begin{equation}
    \begin{aligned}
 & \frac{1}{2}(\mathbf{p} + \mathbf{A}(t))^{2} = \omega - I_{p}, \\
 & \frac{1}{2}(\mathbf{p} + \mathbf{A}(t - \tau))^{2} = - I_{p}.
\end{aligned}
\end{equation}

The joint solutions to the saddle-point equations, Eq.~(S4) for spatial momentum, together with (S6) for time, define the so-called quantum orbits. Each orbit is uniquely characterized by a complex pair of ionization and recombination times \((t_{i},t_{r})\), where
\(t_{r} = t\) and \(t_{i} = t - \tau\). Consequently, the saddle-point canonical momentum \(\mathbf{p}_{s}(t_{i},t_{r})\) depends explicitly on these transition times and takes a different numerical value for each individual quantum orbit.

When analyzing the resulting harmonic spectrum calculated via Eq.~(S5), specific regions can be mapped to the contributions of distinct quantum orbits. In the spectral plateau region, the dominant contributions to the dipole response are provided by the so-called short and long trajectories, which are distinguished by the duration of the electron's excursion in the continuum. For a given harmonic energy, the short trajectory corresponds to a shorter excursion time (typically lower than 0.65 optical cycles), whereas the long trajectory is associated with a more extended propagation. Because these two distinct quantum orbits accumulate different dynamical phases and possess different transverse divergence properties, their coherent superposition generates the characteristic quantum-path interference features observed in HHG. Within our framework, this well-defined pair of trajectories forms the natural basis to isolate a reduced Hilbert subspace, establishing the operational foundation for the APQ.

To separate the short and long quantum-path contributions and construct the operational basis for the APQ defined in the Main Text, we implement a numerical partitioning scheme. While the formal definition of the trajectory-resolved states \(\left| short(t) \right\rangle\) and \(\left| long(t) \right\rangle\) relies on partitioning the continuum integration into distinct momentum volumes \(V_{s}\) and \(V_{s}\), practical numerical evaluations of the full dipole response \(z(t)\) from Eq.~(S1) are more efficiently separated by introducing a smooth temporal filter in the excursion-time domain. To this end, we define a window function

\begin{equation}
\begin{aligned}
    w(\tau,\tau_{\min},\tau_{\max},\Delta\tau) &= \frac{1}{2}\left\lbrack 1 - \tanh\frac{\tau - \tau_{\max}}{\Delta\tau} \right\rbrack\left\lbrack 1 - \frac{1}{2}\left( \tanh\frac{\tau - \tau_{\min}}{\Delta\tau} \right) \right\rbrack,\\ &\quad 0 \leq w \leq 1.
\end{aligned}
\end{equation}

The choice of \(\tau_{short}\) and \(\tau_{long}\), as the characteristic excursion times, can be done either with the semiclassical picture or by means of the saddle-point analysis. Both cases require a proper time--frequency analysis to clearly resolve the separation between the short and long trajectories, as we demonstrate next.

Following the above discussion, it is straightforward to write

\begin{equation}
    z(t) = z_{\short}(t) + z_{\longg}(t).
\end{equation}

The same splitting holds in the frequency domain, yielding
$z(\omega) = z_{\short}(\omega) + z_{\longg}(\omega)$ via Fourier transformation. In Figure~S1 we illustrate the separation method, with a typical SFA-like spectrum for driven by an 8-cycle, 800 nm wavelength pump, as in Eq. (16) in the Main Text . Here, we have set $\tau_{\short} = 0.65$ optical cycles for short trajectories and $\tau_{\longg} = 1.2$ optical cycles for long trajectories (Figure~S1(b))~\cite{Holler2009}. It is straightforward to observe the excellent separation between the short and long trajectory contributions when the harmonic emission is analyzed in a time--frequency representation obtained through a Gabor transform (Figs.~(S1)(c)) and S1(d)). In this representation, the dipole moment is projected onto a localized temporal window, allowing one to resolve the emission time associated with each harmonic frequency. Because short and long trajectories correspond to different excursion times between ionization and recombination, they appear as two well-separated branches in the time-frequency map.

The ability to experimentally and numerically isolate these short and long contributions implies that they can be treated as distinguishable components of the dipole response, which justifies using them as basis states of an effective TLS and APQ description. In this sense, the Gabor analysis does not merely provide a visualization of the semiclassical dynamics but also offers a practical procedure through which the trajectory-defined Hilbert space can be constructed and validated.

The SFA has been demonstrated to be valid in a wide range of strong-field and attosecond applications~\cite{Amini2019}. Because the electron's continuum dynamics are assumed to be driven solely by the external laser field, the approximation most accurately captures the harmonics generated in the plateau region of the spectrum, where the emitted photon energies exceed the ionization potential (\(\omega > I_{p}\))~\cite{Lewenstein1994,Amini2019}. In this regime, the electron typically returns to the core with substantial kinetic energy, and the dominant contribution to the emission is governed by the laser-driven excursion rather than by subtle details of the binding potential. Therefore, the usefulness of the SFA lies not only in its analytical transparency and numerical tractability, but also in the deep physical insight it provides.

\section{Derivation of Eq. (3) of the Main Text}

The temporal separation between the short and long quantum trajectories in the attosecond time-frequency domain (the Gabor map) is a well-established feature of high-harmonic generation. From an operational standpoint, this separation allows one to isolate each trajectory's contribution by applying a time-windowing filter centered around the respective recombination times within the spectral plateau. However, to transition from a signal-processing tool to a rigorous open quantum system framework, this partitioning must be formally mapped onto the electronic Hilbert space, thereby defining the operational two-level subspace of the APQ.

Each quantum orbit in the continuum is uniquely identified by its complex pair of ionization and recombination times, \((t_{i},t_{r})\). Through the saddle-point equations (see Eq.~(S4)), this temporal pair is mapped unambiguously to a specific canonical momentum \(\mathbf{p}_{s}(t_{i},t_{r})\). For a given range of harmonic orders within the spectral plateau, the solutions for \(\mathbf{p}_{s}\) form two distinct, non-overlapping intervals along the laser polarization direction (\(z\)-axis), which we denote as \(\Omega_{s}\) and \(\Omega_{l}\), corresponding to the short and long branches respectively \cite{Schapper2010,Holler2009}. We formally define these intervals as

\begin{equation}
    \Omega_{j} = \left\{ {Re\lbrack p}_{z}\rbrack:Re\left\lbrack p_{s,min}^{j} \right\rbrack \leq {Re\lbrack p}_{z}\rbrack \leq Re\left\lbrack p_{s,max}^{j} \right\rbrack \right\},\ \ j = \left\{ short,long \right\},
\end{equation}
where \(Re\left\lbrack p_{s,min}^{j} \right\rbrack\) (\(Re\left\lbrack p_{s,max}^{j} \right\rbrack\)) represent the minimum and maximum boundaries of the saddle-point canonical momentum for the respective trajectory branch across the spectral plateau. To evaluate these states in the full three-dimensional continuum, the integration domains are defined as the volumes \(V_{j} = \Omega_{j} \times \mathbb{R}^{2}\), where the transverse components span the entire perpendicular plane \((\mathbf{p}_{\perp} \in \mathbb{R}^{2})\). Within the SFA, the continuum electronic wavefunction is typically (but not restricted to, see Ref. \cite{Amini2019}) approximated by Volkov states

\begin{equation}
    \left| \chi_{\mathbf{p}}(t) \right\rangle = \left| \mathbf{p + A}(t) \right\rangle\exp\left( - iS_{\mathbf{p}}(t) \right),
\end{equation}
with \(S_{\mathbf{p}} = \int_{t - \tau}^{t}{dt^{'\frac{1}{2}}}(\mathbf{p} + \mathbf{A}(t'))^{2}\) and \(\left| \mathbf{p + A}(t) \right\rangle\) plane waves. The Volkov states remain mutually orthogonal with respect to their canonical momentum at any given time \(\langle\chi_{\mathbf{p}}|\chi_{\mathbf{p'}}\rangle = \delta(\mathbf{p -}\mathbf{p}^{\mathbf{'}})\). By restricting the momentum integration to the non-overlapping subdomains, we construct two distinct continuum wave packets:

\begin{equation}
    \left| j(t) \right\rangle \equiv \int_{V_{j}}^{}d^{3}\mathbf{p}b\left( \mathbf{p} + \mathbf{A}(t),t \right)\left| \chi_{\mathbf{p}}(t) \right\rangle,\ \ j = \left\{ \short,\longg \right\},
\end{equation}
where \(b\left( \mathbf{p} + \mathbf{A}(t),t \right)\) represents the continuum probability amplitudes derived from the Lewenstein model \cite{Lewenstein1994}. Because the saddle-point contributions of the short and long trajectories are well-separated within the spectral plateau, the overlap between their respective localized momentum distributions is exponentially small. Consequently, this construction ensures that \(|\short(t)\rangle\) and \(|\longg(t)\rangle\) form a highly accurate, approximately orthogonal basis for the reduced continuum subspace at all times.

Filtering these states to the plateau region effectively maps the time-frequency structure of the emission onto the effective Hilbert space of the APQ, as presented in Eq. (3) of the Main Text. Importantly, the continuum wave packets defined in Eq. (S11) encode the quantum information of the entire ensemble of APQs distributed across the spectral plateau. Rather than describing a single, isolated TLS, these time-dependent states represent a collection of coexisting qubits, where each distinct pair of electron paths eventually maps onto a specific harmonic order. Consequently, the time-dependent continuum wavefunction encloses the global electronic wavepacket spreading and retains the sub-cycle history accumulated by the short and long trajectories. Upon recombination, this temporal path-coherence is mapped directly onto the frequency domain, dictating the phase relationships across the harmonic plateau.

It is important to note that the states defined in Eq.~(S11) are not normalized, nor is the electronic state in Eq.~(2) of the Main Text. This lack of normalization arises from the no-depletion approximation, which is fundamentally inconsistent with the transfer of population to the continuum. While this issue is typically neglected when the ionization rate is low, assuming that the majority of the wavefunction remains in the ground state, it directly affects the APQ state. Specifically, the projections onto the short and long subspaces do not yield unit norm. Therefore, the APQ state must be renormalized after projecting it into the trajectory subspace. Although this could be addressed by considering a self-consistently normalized wavefunction, for the scope of this work, we prioritize the physical insight provided by the simplest analytical model to establish the APQ formalism.

\section{Intensity Fluctuations for different Harmonic Orders}

In Figure~S4 we present the Bloch vector trajectory as a function of laser intensity for several harmonic orders considered in this work, as done in the `Dephasing Channel' Section of the Main Text. As the laser intensity varies, it is clear from Figure~S4 that the Bloch vector rotates around the \(z\)-axis. Although all harmonics exhibit this rotation, the behavior is not uniform. Harmonics such as H15 yield a full trip around the \(z\)-axis with minimal variation in the \(z\)-component. In contrast, higher-order harmonics, such as H27, exhibit clear population exchange, and some (such as H51) do not complete a full rotation around the \(z\)-axis. This behavior reflects that each harmonic has an intrinsic, yet slightly different, underlying dynamics when the APQ is defined. Such behavior yields a rich conceptual and experimental framework for investigating the same phenomenon across all available harmonic orders.

\section{One-dimensional Scenario}

All the calculations presented in the Main Text were performed using the full three-dimensional (3D) SFA framework for a linearly polarized electric field, ensuring that quantum wavepacket spreading effects are accounted for. The transverse diffusion is governed by the scaling prefactor \(\left\lbrack 2\pi/(i\tau\  + \epsilon) \right\rbrack^{3/2}\) in Eq. (S3). When a one-dimensional (1D) model is used instead, the transverse degrees of freedom are missing, and the prefactor scales instead as \(\left\lbrack 2\pi/(i\tau\  + \epsilon) \right\rbrack^{1/2}\), meaning that quantum spreading in the perpendicular directions is completely neglected. In the realistic 3D scenario, long trajectories undergo significantly higher transverse spreading than short trajectories simply because they accumulate a longer excursion time \(\tau\) in the continuum, which drastically reduces their recollision probability and suppresses their macroscopic harmonic yield. In contrast, within the 1D approximation, this differential suppression mechanism is diminished, causing the relative weights of the short and long trajectory contributions to become comparable. We illustrate this behavior in Figure~S2, which displays the HHG spectrum calculated under the 1D framework. In Figure~S2(b), it is straightforward to observe that both the short and long trajectory contributions yield relative emission intensities of the same order of magnitude across almost the entire spectral plateau.

To directly quantify how the absence of transverse degrees of freedom impacts the quantum properties of the TLS and the APQ, we display in Figure~S3 the corresponding Bloch vector trajectories as a function of harmonic order and time, together with the interferometric visibility the predictability, and the state purity.

It is clear from Figure~S3(a) that the predictability  \emph{P} lies significantly closer to zero across the plateau compared to the full 3D case presented in Figure~2(a) of the Main Text. This behavior, alongside the modified dynamics of the interferometric visibility \emph{V}, directly reflects how quantum coherence is modulated by the system's dimensionality. Crucially, \emph{V} approaches unity, becoming exactly one for several harmonic orders. This is also reflected in Figure~S3(b), where the coherence map, compared to Figure~2(a) in the Main Text, shows a different spectral evolution with several harmonics reaching the maximum coherence.

This highlights a fundamental physical mechanism about how the dimensionality of the continuum modulates the coherence of the dipole response. In the 3D framework, the laser field couples exclusively to the longitudinal degree of freedom (\(z\)-axis), while the electron transverse momentum components undergo free quantum diffusion. During the continuum excursion, each transverse momentum component \(\mathbf{p}_{\perp}\) accumulates a distinct kinetic phase factor. Because long trajectories spend a longer duration \(\tau\) in the continuum, they accumulate a much wider distribution of these perpendicular phases. When integrating over \(\mathbf{p}_{\perp}\) to evaluate the longitudinal dipole, this phase dispersion leads to a massive destructive. This cancellation is the exact physical origin of the \(\left\lbrack 2\pi/(i\tau\  + \epsilon) \right\rbrack^{3/2}\) damping prefactor, which selectively suppresses the long-trajectory contribution. By restricting the simulation to a 1D scenario, this transverse phase integration is entirely bypassed. Consequently, the phase cancellation is artificially deactivated, preventing the suppression of the long trajectory and locking the interferometric visibility near unity across the plateau.

Furthermore, the population dynamics for the TLS in Figure~S3(c) exhibit a completely different regime from the one shown in Figure~2(c) of the Main Text. Instead of a localized response, the 1D system exhibits highly oscillatory, half-cycle periodic dynamics. This is a direct consequence of the comparable weights of the short and long trajectory contributions in 1D. While the 3D dynamics are strictly restricted to the narrow windows where the short and long trajectories contributions are similar, the equal footing of both branches in 1D enables continuous quantum interference throughout the optical cycle.

In summary, while the one-dimensional model lacks the quantitative accuracy required for a realistic APQ formulation, it remains highly relevant because it isolates the main dynamical mechanisms underlying the core definition. Moreover, it explicitly highlights the role of transverse momentum integration as the dominant decoherence channel, establishing it as a crucial aspect for understanding and controlling the quantum coherence in strong-field attosecond systems.

\section{Transverse Momentum Degree of Freedom: Derivation of Eq.~(33) of the
Main Text}

As established in the Main Text, applying a partial saddle-point approximation exclusively along the laser polarization axis leaves the electron transverse momentum \(\mathbf{p}_{\perp}\) as a conserved continuous label that parametrizes a family of longitudinal quantum paths. To construct the mixed state of the APQ, the unobserved transverse momentum degrees of freedom must be averaged out from the observable response. The continuous part of the electronic Hilbert space is partitioned into longitudinal (parallel to the polarization) and transverse components, such that the continuum states are expanded as

\begin{equation}
    \begin{matrix}
\left| \mathbf{p} \right\rangle = \left| p_{\parallel},\mathbf{p}_{\perp} \right\rangle = \left| p_{\parallel} \right\rangle \otimes |\mathbf{p}_{\perp}\rangle,
\end{matrix}
\end{equation}

Since the quantum trajectories are encoded exclusively within the parallel component, the short-long partitioning of the continuum is restricted to this longitudinal direction, leaving \(\mathbf{p}_{\perp}\) as a purely geometric label. The short-long basis is then defined as

\begin{equation}
    \begin{matrix}
|\short(t,\mathbf{p}_{\perp})\rangle & \equiv & \int_{\Omega_{s}}^{}dp_{z}b(\mathbf{p} + \mathbf{A}(t),t)\left| p_{\parallel}\mathbf{+}A(t) \right\rangle\exp\left( - iS_{\mathbf{p}}(t) \right), \\
|\longg(t,\mathbf{p}_{\perp})\rangle & \equiv & \int_{\Omega_{l}}^{}dp_{z}b(\mathbf{p} + \mathbf{A}(t),t)\left| p_{\parallel}\mathbf{+}A(t) \right\rangle\exp\left( - iS_{\mathbf{p}}(t) \right),
\end{matrix}
\end{equation}
where the integration is carried out only along the polarization axis. The global electronic state is then expanded in terms of this trajectory-resolved basis as

\begin{equation}
    \begin{matrix}
\left| e \right\rangle = e^{iI_{p}t}\left| \phi_{0} \right\rangle + \int_{}^{}{d^{2}\mathbf{p}_{\perp}}(\left| \short\left( t,\mathbf{p}_{\perp} \right) \right\rangle + \left| \longg\left( t,\mathbf{p}_{\perp} \right) \right\rangle)|\mathbf{p}_{\perp}\rangle,
\end{matrix}
\end{equation}
where, again, \(\mathbf{p}_{\perp}\) acts as a continuous label. By projecting this state onto a specific harmonic frequency channel \(q\omega_{0}\), the unnormalized \(\mathbf{p}_{\perp}\)-resolved APQ state is given by

\begin{equation}
\begin{aligned}
    |\psi_{APQ}(q\omega_{0},\mathbf{p}_{\perp})\rangle &\propto \int_{}^{}{d^{2}\mathbf{p}_{\perp}}z_{\short}(q\omega_{0},\mathbf{p}_{\perp})|\short(q\omega_{0})\rangle|\mathbf{p}_{\perp}\rangle \\
    &+ \int_{}^{}{d^{2}\mathbf{p}_{\perp}}z_{\longg}\left( q\omega_{0},\mathbf{p}_{\perp} \right)\left| \longg\left( q\omega_{0} \right) \right\rangle\left| \mathbf{p}_{\perp} \right\rangle,
\end{aligned}
\end{equation}

and therefore, the corresponding density matrix takes the form

\begin{equation}
    \begin{aligned}
    \rho_{APQ}\left( q\omega_{0},\mathbf{p}_{\perp} \right) &\propto \iint_{}^{}{d^{2}\mathbf{p}_{\perp}}d^{2}\mathbf{p'}_{\perp}
    {|z_{\short}|}^{2}\left| \short \right\rangle\left\langle \short \right|\left| \mathbf{p}_{\perp} \right\rangle\langle\mathbf{p'}_{\perp}| \\
    &+ \iint_{}^{}{d^{2}\mathbf{p}_{\perp}}d^{2}\mathbf{p'}_{\perp}{|z_{\longg}|}^{2}\left| \longg \right\rangle\left\langle \longg \right|\left| \mathbf{p}_{\perp} \right\rangle\langle\mathbf{p'}_{\perp}| \\
    &+ \iint_{}^{}{d^{2}\mathbf{p}_{\perp}}d^{2}\mathbf{p'}_{\perp}z_{\short}z_{\longg}^{*}\left| \short \right\rangle\left\langle \longg \right|\left| \mathbf{p}_{\perp} \right\rangle\langle\mathbf{p'}_{\perp}| \\
    &+ \iint_{}^{}{d^{2}\mathbf{p}_{\perp}}d^{2}\mathbf{p'}_{\perp}z_{\longg}z_{\short}^{*}\left| \longg \right\rangle\left\langle \short \right|\left| \mathbf{p}_{\perp} \right\rangle\langle\mathbf{p'}_{\perp}|,
    \end{aligned}
\end{equation}
where we have suppressed the explicit \(q\omega_{0}\) dependence in the short and long states for sake of simplicity. Note that because of the continuous transverse momentum states \(\left| \mathbf{p}_{\perp} \right\rangle\), this global density matrix is formally infinite-dimensional. To isolate the discrete two-level subspace of the APQ, we perform a partial trace over these unobserved transverse degrees of freedom

\begin{equation}
    \begin{aligned}
{{Tr}_{\mathbf{p}_{\perp}}\lbrack\rho}_{APQ}(\omega,\mathbf{p}_{\perp}) & \rbrack = 
\end{aligned}\int_{}^{}{d^{2}\mathbf{p}_{\perp}}\left\langle \mathbf{p}_{\perp}\left| \rho_{APQ}\left( \omega,p_{\perp} \right) \right|\mathbf{p}_{\perp} \right\rangle.
\end{equation}

Evaluating this partial trace yields the reduced, unnormalized APQ density matrix

\begin{equation}
    \begin{aligned}
\rho_{APQ}(\omega) & \propto \int d^{2}p_{\perp}\begin{pmatrix}
|z_{\short}(\omega,p_{\perp})|^{2} & & z_{\short}(\omega,p_{\perp})z_{\longg}^{*}(\omega,p_{\perp}) \\
z_{\short}^{*}(\omega,p_{\perp})z_{\longg}(\omega,p_{\perp}) & & |z_{\longg}(\omega,p_{\perp})|^{2}
\end{pmatrix}
\end{aligned},
\end{equation}
where the integration is understood to be carried out element by element. Finally, to ensure a valid statistical interpretation for the trajectory subspace, the final state is locally normalized via the same physical principles described in the previous sections.

From the numerical evaluation of the saddle-point equations, and using the laser parameters introduced in Eq. (16) of the Main Text, the parallel momentum \({|p}_{z}|\) ranges between 0.06 a.u. and 0.65 a.u. for the long trajectories, whereas for the short trajectories it typically lies between 0.64 a.u. and 1.85 a.u. near the cutoff region. A similar consideration applies to the transverse momentum integration bounds. In principle, the partial trace over the perpendicular degrees of freedom defined in Eq. (S18) requires integrating over the entire infinite two-dimensional space (\(\mathbf{p}_{\bot} \in \ \mathbb{R}^{2}\ \ \)). However, due to the exponential suppression inherent to both the tunnel ionization rate and the subsequent transverse quantum wavepacket spreading, the sub-cycle dipole amplitudes \(z_{j}(\omega,\ \mathbf{p}_{\bot}\)) decay rapidly as a function of the perpendicular momentum. As demonstrated by the numerical convergence analysis illustrated in Figure~6 of the Main Text, evaluating the integration within a bounded domain of \(\mathbf{p}_{\bot}| \leq 1\) a.u. is more than sufficient to fully capture the macroscopically observed coherence properties and state purity, ensuring strict numerical accuracy.

\bibliography{refs}

\newpage

\begin{figure}[ht]
    \centering
    \includegraphics[width=0.9\linewidth]{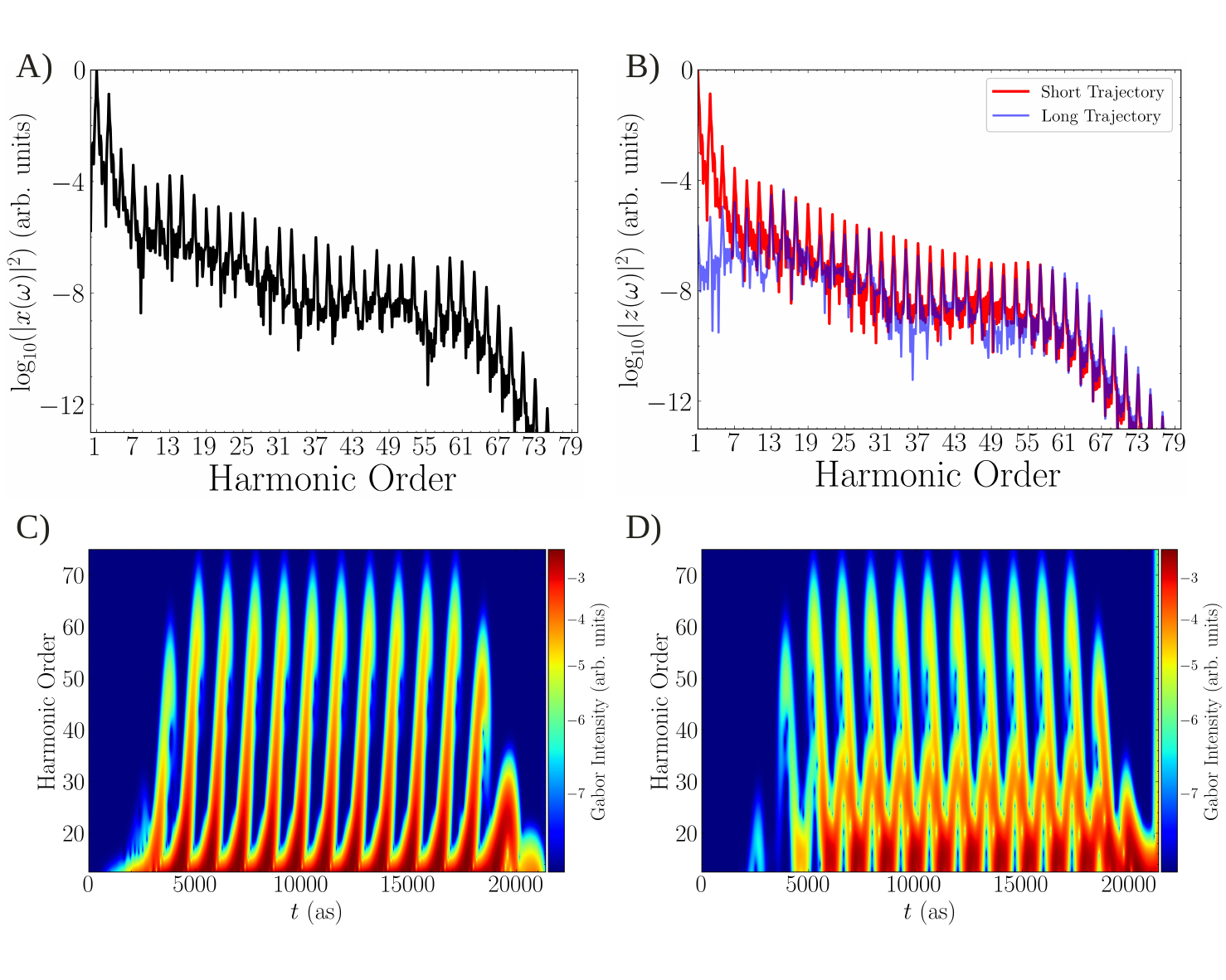}
    \caption{\textbf{HHG spectra for a Hydrogen target.} The driven field is an 8-cycle trapezoidal-shaped laser pump, with a wavelength \(\lambda = 800\) nm and intensity \(I = 4.2 \times 10^{14}\) W/cm\(^{2}\). (A) Normalized total spectra, and (B) short and long trajectory-based spectra calculated with the window defined in Eq.~(S7). (C) Time-frequency analysis for the short trajectory-based spectra, and (D) for the long trajectory-based spectra.}
    \label{fig:S1}
\end{figure}

\begin{figure}[ht]
    \centering
    \includegraphics[width=0.9\linewidth]{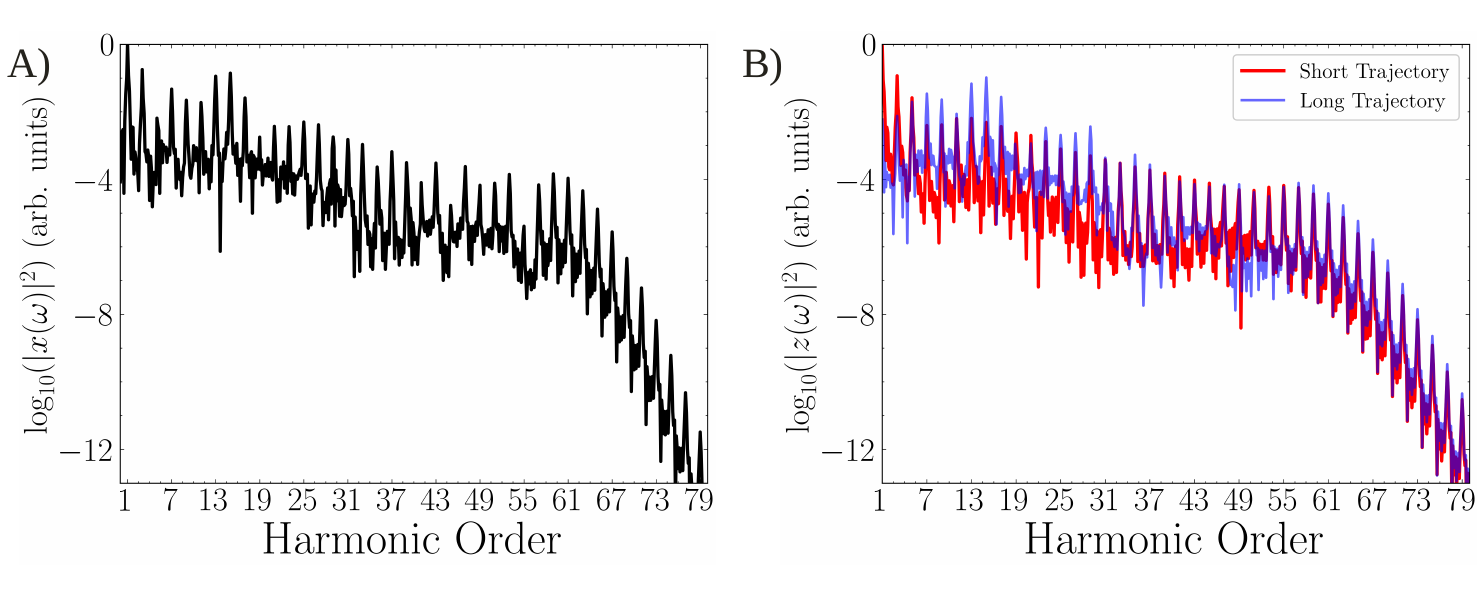}
    \caption{\textbf{HHG spectra for a Hydrogen target when the simulations are one-dimensional.} The driven field is an 8-cycle trapezoidal-shaped laser pump, with a wavelength \(\lambda = 800\) nm and intensity \(I = 4.2 \times 10^{14}\) W/cm\(^{2}\) (Eq. (16) in the Main Text). (A) Normalized total spectra, and (B) short and long trajectory-based spectra calculated with the window defined in Eq.~(S7).}
    \label{fig:S2}
\end{figure}

\begin{figure}[ht]
    \centering
    \includegraphics[width=0.9\linewidth]{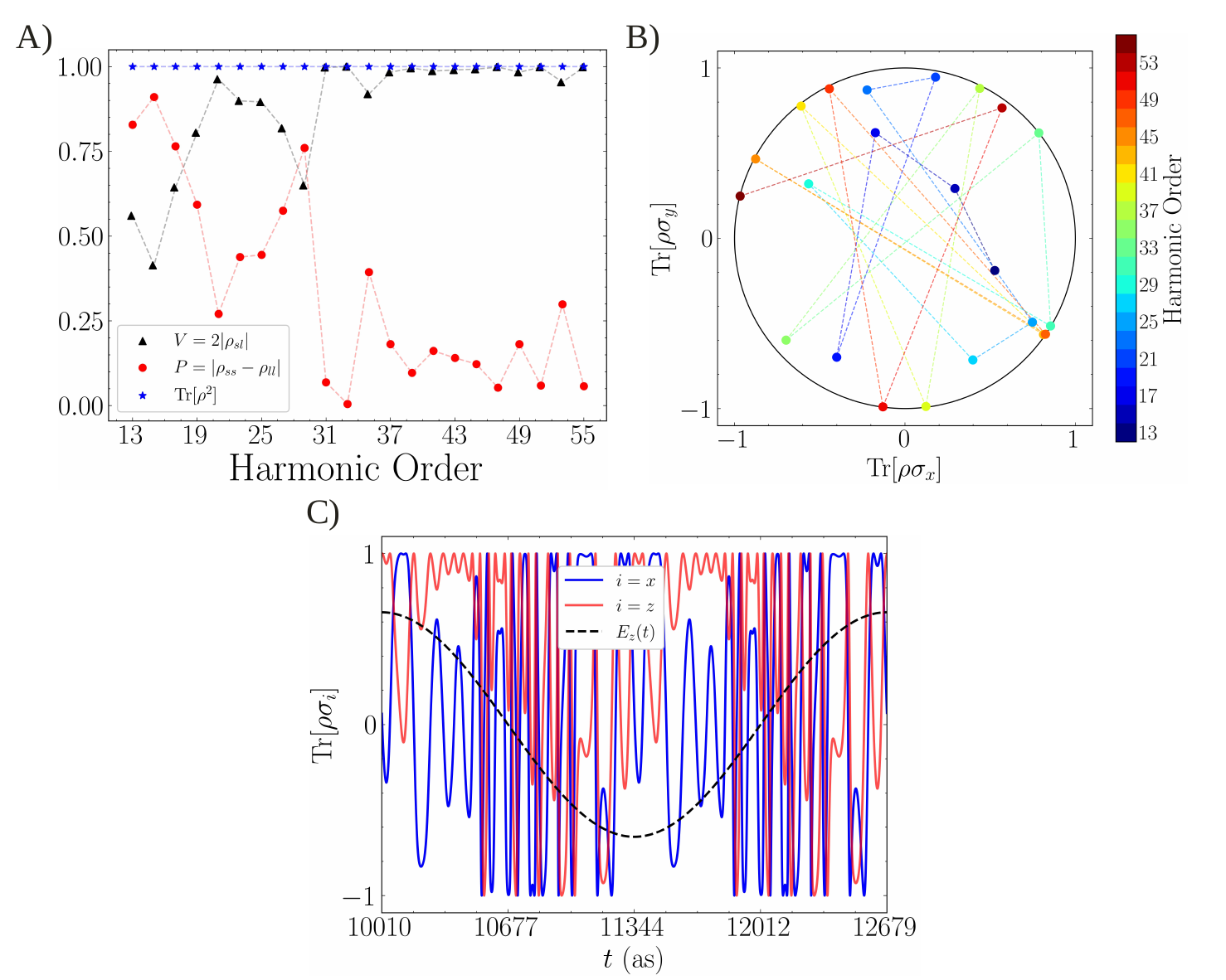}
    \caption{\textbf{APQ and TLS Bloch vector dynamics for pure-state density matrices}. (A) Interferometric visibility \emph{V} (black triangles), predictability \emph{P} (red circles), and state purity \(\mathbf{Tr\lbrack}\mathbf{\rho}^{\mathbf{2}}\mathbf{\rbrack}\) (blue stars) as functions of harmonic order for the APQ. (B) Complex coherence map showing the \(\mathbf{x}\) and \(\mathbf{y}\) Bloch-vector components as a function of harmonic order; comparing with Figure~2(b) in the Main Text, the expansion towards the disk boundary illustrates how coherence is modulated by the system's dimensionality. (C) Real-time evolution of the Bloch-vector components during one optical cycle for the time-domain TLS, for the driving laser field defined in Eq. (16) in the Main Text.}
    \label{fig:S3}
\end{figure}

\begin{figure}[ht]
    \centering
    \includegraphics[width=0.9\linewidth]{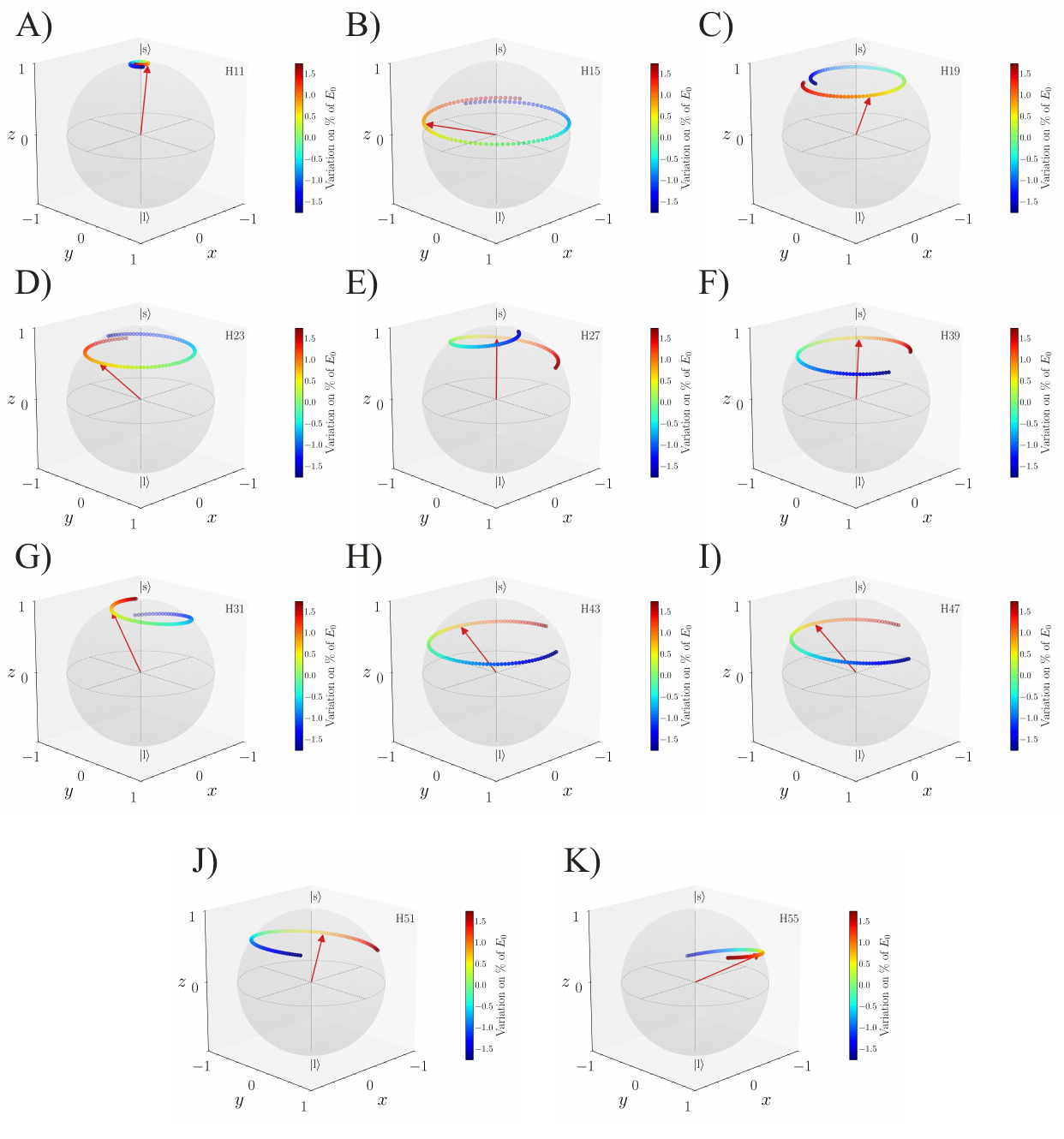}
    \caption{ \textbf{APQ Bloch vector trajectory for several harmonic orders between H11 and H55 (A-K, see panel labels), as a function of the laser intensity.} The scale is presented as a percentage of variation from the central value, \(E_{0} = 0.109\) a.u.}
    \label{fig:S4}
\end{figure}